\documentclass[aps,prmaterials,reprint,superscriptaddress,nofootinbib,amsmath,amssymb]{revtex4-2}

\usepackage{times}
\usepackage{graphicx}
\usepackage{booktabs}
\usepackage{gensymb}
\usepackage{xcolor}
\usepackage[colorlinks]{hyperref}

\newcommand{\hafnia}{HfO$_2$}
\newcommand{\silica}{SiO$_2$}
\newcommand{\sdh}{SiO$_2$-doped HfO$_2$}
\newcommand{\tantala}{Ta$_2$O$_5$}
\newcommand{\titania}{TiO$_2$}

\begin{document}

\preprint{APS/123-QED}

\title{Network Topology of Hafnia-Based Amorphous Optical Coatings by Grazing-Incidence X-ray Total Scattering and Atomic Modeling}

\author{K. Prasai}
\email{kprasai@kennesaw.edu}
\affiliation{Department of Physics, Kennesaw State University, Marietta, Georgia 30060, USA}

\author{B. LaBell}
\affiliation{Department of Physics, Kennesaw State University, Marietta, Georgia 30060, USA}

\author{Kyung-ha Lee}
\affiliation{Department of Physics, Sungkyunkwan University, Seoul 03063, Republic of Korea}

\author{A.\,Mehta}
\affiliation{SLAC National Accelerator Laboratory, Menlo Park, California 94025, USA}

\author{B.\,Shyam}
\affiliation{Xerion Advanced Battery Corp., Kettering, Ohio 45420, USA}

\author{M.\,M.\,Fejer}
\affiliation{E. L. Ginzton Laboratory, Stanford University, Stanford, California 94305, USA}
                      
\author{R.\,Bassiri}
\affiliation{E. L. Ginzton Laboratory, Stanford University, Stanford, California 94305, USA}

\date{August 10, 2026}

\begin{abstract}
\noindent Amorphous hafnia-based films are promising optical-coating materials for cryogenic GW detectors, but their performance depends on how doping and annealing modify the atomic network. We combine grazing-incidence X-ray total scattering measurements with experimentally constrained atomic modeling to study the as-deposited \hafnia\ and 27\% \sdh\ films annealed at 150$^\circ$C and 400$^\circ$C. Pure amorphous \hafnia\ is a dense, high-coordination network of Hf-centered polyhedra with substantial edge- and face-sharing connectivity. Incorporating \silica\ introduces stable SiO$_4$ tetrahedra, lowers the Hf and O coordination, and replaces highly connected Hf-rich oxygen environments with mixed Si--O--Hf bridges. This produces a chemically mixed network rather than isolated \silica-rich regions, and shifts the cation topology toward corner-sharing connectivity. Annealing to 400$^\circ$C produces only modest additional structural relaxation. These results provide an atomic-scale description of how \silica\ modifies the topology of amorphous \hafnia-based coatings and suggest structural descriptors relevant to understanding their mechanical-loss behavior.
\end{abstract}

\pacs{Valid PACS appear here}

\maketitle

\section{INTRODUCTION}\label{sec:introduction}

The sensitivity of interferometric gravitational-wave detectors, including Advanced LIGO and Virgo, is limited by several fundamental and technical noise sources. An important source of noise, coating thermal noise (CTN), originating from the mechanical loss of the mirror coatings used to form the high-reflectivity test-mass mirrors, directly limits the detector sensitivity \cite{martynov2016sensitivity,buikema2020sensitivity}. The current generation of gravitational-wave detectors uses dielectric Bragg mirrors consisting of alternating layers of ion-beam-sputtered amorphous high- and low-refractive-index oxides, \titania-doped \tantala\ and \silica\ \cite{harry2006titania,steinlechner2018development,amato2019optical,granata2020amorphous}. Planned upgrades and future observatories, including Advanced LIGO+, Cosmic Explorer, and Einstein Telescope, require further reduction of coating thermal noise while maintaining low optical absorption, low scatter, and high reflectivity \cite{miller2015prospects,AplusWP,punturo2010einstein,hild2012beyond,hall2021gravitational,CEcoatingsWP}. Identifying new amorphous coating materials and understanding how their atomic structure controls mechanical and optical properties are therefore central problems in the development of next-generation precision optical coatings.

Hafnia-based coatings have been considered for low-thermal-noise coatings for future gravitational-wave detectors \cite{abernathy2011cryogenic, bassiri2011atomic}. 
Amorphous or glassy \hafnia\ is generally regarded as a poor glass former \cite{abernathy2011cryogenic, hill2008relationship} and is structurally difficult to isolate, with crystallization occurring readily during deposition or subsequent thermal treatment. Incorporating \silica\ can suppress \hafnia-like crystalline order and improve amorphous-phase stability, although the extent of this stabilization depends strongly on composition and thermal history \cite{ushakov2004crystallization,hill2008relationship,afify2006xrd}. For cryogenic detectors, 27\% \sdh\ has been proposed as a low-index partner for amorphous silicon in multimaterial coating designs, reducing thermal noise relative to conventional \silica-containing stacks while meeting stringent optical-absorption requirements \cite{Craig2019PRL}.

Mechanical loss in amorphous oxides is commonly described in terms of two-level systems (TLS) associated with local structural rearrangements in the disordered network. Although the microscopic origin of TLS is not unique, atomistic studies increasingly indicate that mechanical dissipation is associated with specific short- and intermediate-range structural motifs. Edge- and face-sharing polyhedra, for example, have been shown to  correlate with room-temperature  mechanical loss \cite{prasai2019high, prasai-submitted}, whereas lower-barrier and more localized rearrangements particularly oxygen-centered, become important at cryogenic temperatures \cite{puosi2020nonlocal, damart2018atomistic}. Mechanical loss therefore depends not only on composition, but also on network topology, including cation coordination, bond-angle distributions, polyhedral connectivity, and intermediate-range order. These structural features are themselves sensitive to deposition conditions, annealing, and dopant concentration \cite{vajente2018effect,fazio2022comprehensive,prasai2019high, prasai-submitted}, underscoring the need to establish quantitative relationships between processing, atomic structure, and mechanical dissipation. 

Earlier EXAFS studies showed that amorphous \hafnia\ contains locally coordinated Hf--O environments that cannot be described simply by any single crystalline phase \cite{Cho2008HfO2EXAFS,Haque2016HfO2EXAFS,Afify2006SiO2HfO2EXAFS,Liu2024HfO2Orthorhombic,Ozkendir2025HfO2XAFS}. First-principles and scattering studies have shown amorphous \hafnia\ exhibiting broad distributions of coordination environments and bond angles characteristic of a structurally disordered oxide network \cite{Nguyen2019AmorphousHfO2,Chen2011AmorphousHfO2,Kumar2021HfO2Optoelectronic,Perevalov2007HfO2,Zhao2002HfO2}.
It has also been shown that Hf coordination and HfO$_n$ polyhedral connectivity are sensitive to density and thermal history \cite{gallington2017structure,nguyen2019molecular}. Incorporation of \silica\ further modifies this network by introducing tetrahedral SiO$_4$ units among the more highly coordinated Hf-centered polyhedra, creating a competition between relatively open and compact structural motifs. How this competition modifies Hf coordination, Hf--O--Hf and Si--O--Hf connectivity, intermediate-range order, and their evolution upon annealing remains poorly understood in ion-beam-sputtered optical coatings. Resolving these structural changes requires measurements and atomistic models capable of determining local coordination, bond-angle distributions, polyhedral connectivity, and intermediate-range organization. 

Grazing-incidence X-ray pair distribution function (GIPDF) measurements provide a direct way to probe the short- and intermediate-range order of amorphous thin films while eliminating any interference from the substrate. Thin-film PDF and GIPDF methods have previously been used to measure local structure in amorphous and crystalline films and to study amorphous optical coatings such as \tantala-based materials \cite{jensen2015demonstration,shyam2016measurement,prasai2019high}. The measured PDF contains information about interatomic distances and medium-range correlations, but interpretation of the total PDF requires atomic models because different partial pair correlations overlap strongly in $r$ space. This is especially true for \hafnia-based films, where the large X-ray scattering weight of Hf strongly emphasizes Hf-containing correlations. Atomic modeling is therefore needed to separate Hf--O, Hf--Hf, Si--O, Si--Hf, and O--O contributions and to translate the measured GIPDF into more detailed network descriptors such as coordinations, bond angles and polyhedral-connectivity.

In this paper, we report a combined GIPDF and atomistic-modeling study of amorphous \hafnia\ and \sdh\ optical coatings. We study three films: as-deposited pure \hafnia, 27\% \sdh\ annealed at 150$^\circ$C, and 27\% \sdh\ annealed at 400$^\circ$C. For each sample, we measure the GIPDF and construct an ensemble of atomic models by a combination of fitting structure to the measured data and energy minimization using a machine-learning-based forcefield. We then use the models to analyze the structural differences between pure \hafnia\ and \sdh, and between the \sdh\ films annealed at 150$^\circ$C and 400$^\circ$C. The goal of this work is to determine how \silica\ incorporation and annealing modify the Hf--O network, Hf coordination, polyhedral packing, and intermediate-range order of amorphous \hafnia-based optical coatings. These results provide the atomic-structure foundation needed to understand why \sdh\ is promising for low-noise optical coatings and how its structure might be further optimized for future gravitational-wave detectors.

\section{METHODS} \label{sec:methods}

\subsection{Film Deposition and Characterization}

The \hafnia\ and \sdh\ films studied in this work were deposited by ion-beam sputtering (IBS), following the coating preparation described in Ref.~\cite{Craig2019PRL}. For the mechanical-loss measurements reported in the reference, the coatings were deposited on crystalline-Si cantilevers. The \sdh\ coating contained 27\% \silica, as measured by x-ray photoelectron spectroscopy (XPS). The as-deposited thickness of the \sdh\ coating was measured by ellipsometry to be $483\pm3$~nm.

Companion fused-silica substrates were coated in the same deposition run for optical and structural characterization. The samples used in the present grazing-incidence x-ray scattering measurements were taken from this same coating set. For the \sdh\ films, post-deposition heat treatments were carried out for 24~h at temperatures of 150$^\circ$C and 400$^\circ$C. Ellipsometry measurements showed no significant change in the thickness of the \sdh\ coating after heat treatment. Transmission electron microscopy measurements on coatings deposited on \silica\ substrates showed that the heat-treated \sdh\ films remained amorphous up to 600$^\circ$C.

\subsection{GIPDF Measurements}
GIPDF data were collected at the dedicated X-ray scattering beamline 10-2 at the Stanford Synchrotron Radiation Lightsource (SSRL). Using an energy-resolved point detector to scan over scattering angles, a $q$-dependent scattering profile was collected for each sample. A grazing-incidence angle was chosen to preferentially collect X-rays scattered from the coatings (rather than the substrate) while maintaining a $q$-range up to 20.1\,\AA$^{-1}$. The X-ray energy was 21.5~keV. The elastic signal from scattered X-rays was used to obtain the total scattering signal from the coating, while the fluorescence signal from the sample was used to correct for the detector footprint in the measured intensities. The total scattering data were reduced to the normalized structure factor $S(q)$ after applying identical corrections to all samples for air scattering, absorption, Compton scattering, and polarization effects. The structure factors were then Fourier-transformed to $r$-space to obtain the X-ray PDFs for each sample. Further details of the GIPDF data-collection method are discussed in Refs.~\cite{shyam2016measurement,qiu2004pdfgetx2}.

\subsection{Density determination}
\label{sec:density}

\begin{figure*}[t]
	\centering
	\includegraphics[width=\linewidth]{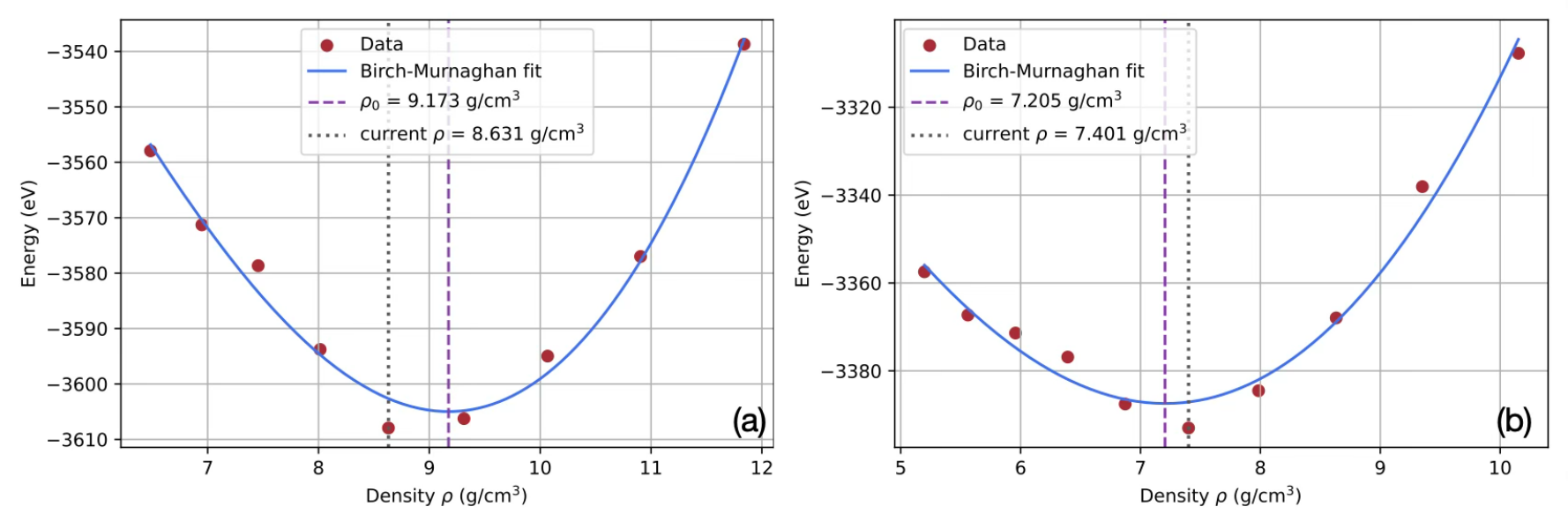}
\caption{Density determination from Birch--Murnaghan fits to the energy--volume data \cite{birch1947finite}. Total energies from DFT-relaxed structures are shown as red circles. For clarity, the horizontal axis is shown as the corresponding mass density, $\rho$, rather than cell volume. Panel (a) shows pure \hafnia\ for which the fitted equilibrium density is $\rho_0=9.17$~g/cm$^3$. Panel (b) shows 27\% \sdh\ for which the fitted equilibrium density is $\rho_0=7.21$~g/cm$^3$. The black dotted lines mark the densities of the original melt--quench models before equation-of-state fitting, while the purple dashed lines mark the fitted equilibrium densities used for the GIPDF-constrained structural modeling.}
\label{fig:density_eos}
\end{figure*}

The mass density of HfO$_2$ films depends on the deposition method and processing conditions, with experimental studies reporting densities around 9~g/cm$^{3}$ for amorphous films \cite{modreanu2005solid,puurunen2005hafnium}. In order to guide the modeling effort, it would be ideal to measure the density of the films on which PDF measurement have been carried out. However, for various reasons, we could not get the density measured independently for the exact samples studied here. As an alternative way, we numerically estimated the density by generating melt-quench based models and performing an equation of state fit on the volume--energy data. 

For each composition, an amorphous structure was first generated by melt--quench molecular dynamics using the MACE machine-learning interatomic potential \cite{batatia2022mace}. The pure \hafnia\ model contained Hf$_{120}$O$_{240}$, and the 27\% \sdh\ model contained Hf$_{88}$Si$_{32}$O$_{240}$, corresponding to a Si cation fraction of 26.7\%. Each melt--quenched configuration was uniformly rescaled to a series of trial cell volumes. At each fixed volume, the atomic positions were relaxed using density-functional theory (DFT) while keeping the lattice vectors fixed. The DFT calculations were carried out using VASP \cite{kresse1996efficient}, the PBE exchange-correlation functional \cite{perdew1996generalized}, and PAW potentials \cite{blochl1994projector}. A plane-wave cutoff of 400~eV and a $\Gamma$-point $1\times1\times1$ $k$-point mesh were used. Ionic relaxations were performed until the force convergence criterion of $10^{-2}$~eV/\AA\ was reached. A static calculation was then performed at each volume.

The resulting energy--volume data were fit using the third-order Birch--Murnaghan equation of state \cite{birch1947finite}. The fitted equilibrium volume was converted to the corresponding mass density. The densities obtained from these fits are summarized in Table~\ref{tab:density_eos}. The fitted density for pure \hafnia\ was 9.17~g/cm$^3$, while that for 27\% \sdh\ was 7.21~g/cm$^3$. These values were used for the GIPDF-constrained structural modeling. Because the two \sdh\ samples have the same nominal composition, the same fitted density was used for both the 150$^\circ$C and 400$^\circ$C \sdh\ models.

\begin{table}[t]
\caption{Computed mass densities used for the structural modeling. The quoted uncertainties are the formal one-standard-error values obtained from the nonlinear least-squares Birch--Murnaghan fits. The covariance matrix of the fitted parameters was used to estimate the uncertainty in the equilibrium volume $V_0$, which was then propagated to the density through $\rho=M_{\rm cell}/V_0$.}
\label{tab:density_eos}
\begin{ruledtabular}
\begin{tabular}{lc}
System & Computed density \\
 & (g/cm$^3$) \\
\hline
\hafnia & $9.17 \pm 0.11$ \\
27\% \sdh & $7.21 \pm 0.14$ \\
\end{tabular}
\end{ruledtabular}
\end{table}

\subsection{Generating Structure Models}
\label{sec:generating_models}
Atomic models were generated for all three measured GIPDF datasets: as-deposited pure \hafnia, 27\% \sdh\ heat treated at 150$^\circ$C, and 27\% \sdh\ heat treated at 400$^\circ$C. For pure \hafnia, the simulation cell contained 2880 atoms and the fitted equilibrium density from Sec.~\ref{sec:density}, $\rho_0=9.17$~g/cm$^3$, was used. For 27\% \sdh models, the simulation cell contained 2880 atoms and the fitted density $\rho_0=7.21$~g/cm$^3$ was used for both the 150$^\circ$C and 400$^\circ$C datasets. For each dataset, ten independent amorphous starting configurations of size 360 atoms were generated by melt--quench molecular dynamics using MACE machine-learning potential. Then, the starting configurations, of size 2880 atoms, were prepared by stacking these melt-quench MD models in 2$\times$2$\times$2 grid following the `building-block' method \cite{drabold2009topics}. These starting structures were then used as the initial configurations for the GIPDF-constrained refinement described below.

Developing reliable models of amorphous thin films from diffraction data, such as GIPDF, is a nontrivial inverse problem. Reverse Monte Carlo (RMC) and related regression methods can fit the measured pair correlations, but unaided coordinate fitting can produce nonphysical structures because the diffraction data do not uniquely determine the full three-dimensional atomic configuration \cite{keen1990structural,mcgreevy2001reverse}. A more reliable strategy is to combine experimental fitting with energetic constraints, so that the resulting models reproduce the measured PDF while remaining chemically and mechanically reasonable. Here we use an iterative refinement procedure closely related to the Force Enhanced Atomic Refinement (FEAR) method \cite{pandey2015force,pandey2016inversion}. In each refinement cycle, RMC moves were first used to reduce $\chi^2$, the difference between the measured and calculated structure defined as 

\begin{equation}
\chi^2 =
\sum_j
\frac{
\left[
S^{\rm expt}(q_j)-S^{\rm calc}(q_j)
\right]^2
}{
\sigma_j^2
},
\label{eq:chi2}
\end{equation}
where the sum is over discrete $q$ bins, $S^{\rm expt}(q)$ and $S^{\rm calc}(q)$ are the experimental and calculated structure factors, respectively, and $\sigma_j$ is a weighting factor. In this work, we set $\sigma_j=1.0$ for all $q$. After a fixed number of accepted RMC moves, few steps of energy minimization were carried out using the MACE machine-learning interatomic potential \cite{batatia2022mace} and the minimization algorithm of Ref. \cite{bitzek2006structural}. These RMC and energy minimization moves were incomplete in the sense that only a limited number of moves was performed for each before switching to the other. In particular implementation for this paper, 100 steps of RMC accepted moves are alternated with 5 steps of energy-minimization steps. A bond-valence-sum penalty was included in the effective RMC cost function to discourage chemically unreasonable coordination environments \cite{norberg2009bond}. The iterative switching between RMC and energy minimization was repeated until both $\chi^2$ and the MACE energy approached a plateau.

\begin{figure}[t]
	\centering
\includegraphics[width=\linewidth]{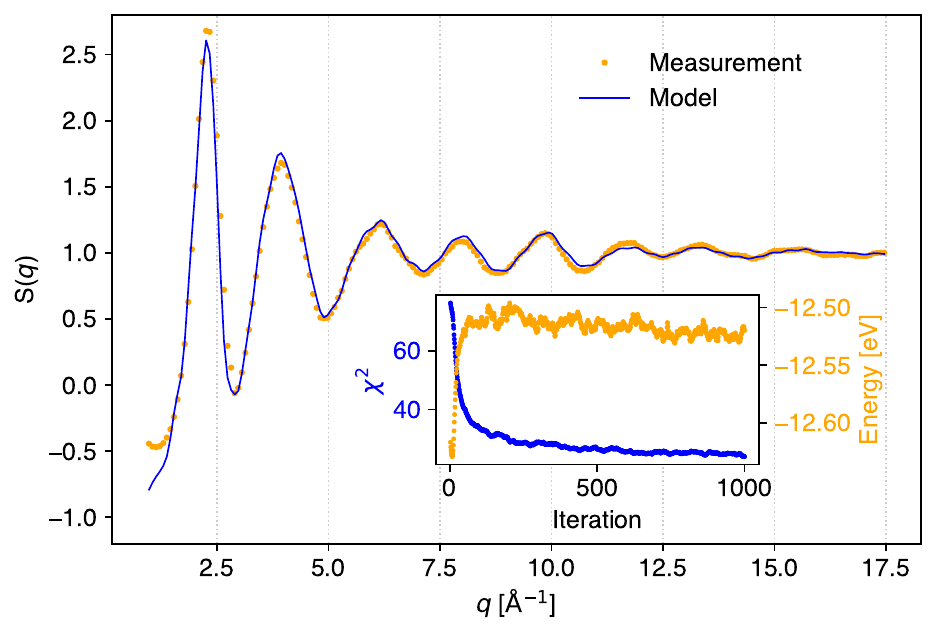}
	\caption{Example of GIPDF-constrained structural refinement for one \hafnia-based model. The main panel compares the measured and modeled structure factors after the final refinement cycle. The insets show the evolution of $\chi^2$ and the average MACE energy during the iterative RMC and energy-minimization procedure.}
	\label{fig:rmc_convergence}
\end{figure}

The final models, by construction, agree well with the measured x-ray data; an example is shown in Fig.~\ref{fig:rmc_convergence}. The decrease in $\chi^2$ is strongest during the early refinement cycles and then slows as the model approaches convergence. The energy does not necessarily decrease monotonically over the full refinement because the procedure minimizes an effective cost function that includes both agreement with experiment and energetic regularization. Thus, the final structure may move away from the local MACE minimum when required to reproduce experimentally constrained pair correlations. This should not be interpreted as a loss of physicality, but as the cost of imposing structural information contained in the GIPDF data that is not fully captured by the starting melt--quench structures alone.

For each sample, the full refinement was carried out for ten independent starting configurations and random seeds. In the remainder of this paper, each set of ten final configurations is referred to as the model ensemble for the corresponding sample. All reported structural quantities are averages over these ten models, and the error bars represent the corresponding standard deviations.

\section{RESULTS AND DISCUSSION}\label{sec:results}

\subsection{Measured GIPDFs and validity of the atomic models}

\begin{figure}[t]
	\centering
    \includegraphics[width=\linewidth]{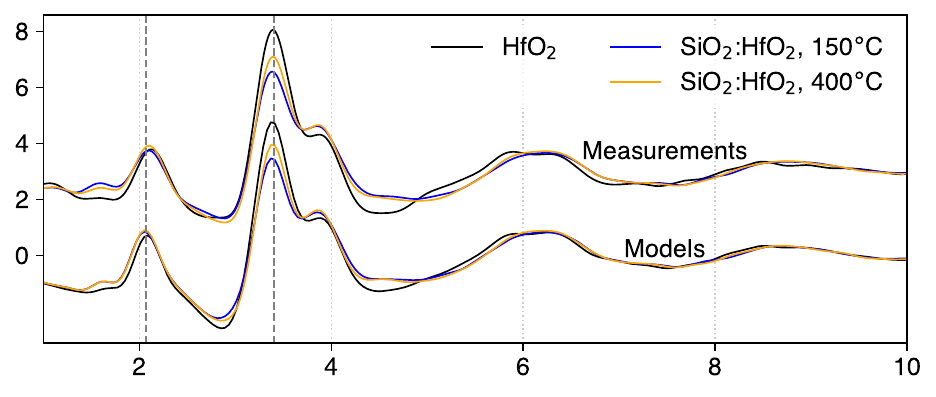}
	\caption{Measured and modeled PDFs of amorphous \hafnia\ and 27\% \sdh\ films. The main peaks near 2.07~\AA\ and 3.40~\AA\ are indicated by dashed lines. The modeled PDFs reproduce the measured peak positions and the main changes between pure \hafnia\ and \sdh. The PDFs are shifted vertically for clarity.}
	\label{fig:gipdf_hfo2_sdh}
\end{figure}

The measured GIPDFs of the three films are shown in Fig.~\ref{fig:gipdf_hfo2_sdh}. All three curves exhibit broad peaks and no long-range crystalline oscillations, consistent with amorphous oxide coatings. The as-deposited pure-\hafnia\ film shows a strong first peak near 2.1~\AA, followed by a strong cation--cation peak near 3.4~\AA. The first peak is assigned mainly to nearest-neighbor Hf--O correlations, while the second peak is dominated by Hf--Hf correlations. The 27\% \sdh\ films show the same main Hf-containing features but with modified peak shapes because part of the Hf--O network is replaced by Si-centered tetrahedra. In addition, \sdh\ models have a prepeak near 1.6~\AA~that corresponds to SiO$_4$ tetrahedra (see discussion in Sec. \ref{sec:results_partials}).

The modeled PDFs reproduce the positions of the main experimental peaks in all three samples. In particular, the agreement in peak positions and in the relative changes between the samples indicates that the models capture the main short- and intermediate-range structural features needed for the analysis below.

The difference between the two \sdh\ measurements is comparatively small. The 400$^\circ$C annealed film has a slightly sharper first-shell peak and a slightly modified cation--cation region compared with the 150$^\circ$C film. These changes are much smaller than the difference between pure \hafnia\ and \sdh. Thus, incorporation of \silica\ produces the dominant structural modification, while annealing mainly relaxes the mixed network without changing its basic local topology. This conclusion is supported by the coordination and polyhedral-connectivity statistics discussed below.

\subsection{Partial pair correlations and oxygen-bridge analysis} \label{sec:results_partials}
\begin{figure}[t]
	\centering
	\includegraphics[width=\linewidth]{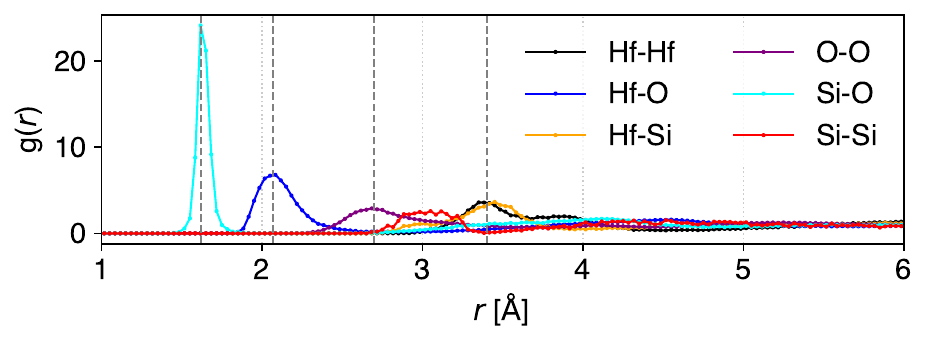}
	\caption{Number-density-based partial PDFs (defined following Eq.~8 of Ref.~\cite{keen2001comparison}) from the atomic models of 400$^\circ$C-annealed \sdh\ models. Dashed lines are drawn at 1.62 \AA, 2.07 \AA, 2.70 \AA, and 3.40 \AA\ as guidance to locate major peak positions.}
	\label{fig:partial_pdfs_hfo2_sdh}
\end{figure}

To identify the structural origin of the measured PDF features, we computed number-density-based partial PDFs from the atomic models, and those from 400$^\circ$C-annealed \sdh\ models are shown in Fig.~\ref{fig:partial_pdfs_hfo2_sdh}. The first peak in the total PDF is dominated by Hf--O correlations centered near 2.07~\AA. The O--O partial PDF has its main first-neighbor feature near 2.7~\AA, but this feature is not noticeable in the measured x-ray PDF because oxygen has much smaller x-ray scattering weight compared with Hf. The second peak in the

\begin{table*}[t]
\caption{Oxygen-bridge environments from the final atomic models. For each oxygen atom, neighboring cations were identified using Hf--O and Si--O cutoffs of 2.8~\AA\ and 2.2~\AA, respectively. Values are percentages of oxygen atoms, averaged over the ten models in each ensemble. The quoted uncertainties are model-to-model standard deviations, reflecting the structural variation among independently refined models. The ``Other'' category contains rare environments with populations below the main listed categories.}
\label{tab:oxygen_bridge}
\begin{ruledtabular}
\begin{tabular}{lccc}
Oxygen environment &
\hafnia, AD &
27\% \sdh, 150$^\circ$C &
27\% \sdh, 400$^\circ$C \\
\hline
Hf--O--Hf       & $5.18 \pm 1.20$  & $7.37 \pm 1.05$  & $7.25 \pm 1.09$ \\
Hf--O--Si       & $0.00 \pm 0.00$  & $16.80 \pm 2.22$ & $17.29 \pm 2.21$ \\
Si--O--Si       & $0.00 \pm 0.00$  & $3.81 \pm 1.18$  & $3.89 \pm 1.18$ \\
Hf$_3$--O       & $67.43 \pm 2.84$ & $37.89 \pm 1.98$ & $38.24 \pm 1.73$ \\
Hf$_2$Si--O     & $0.00 \pm 0.00$  & $24.11 \pm 1.25$ & $23.78 \pm 1.05$ \\
HfSi$_2$--O     & $0.00 \pm 0.00$  & $1.60 \pm 0.74$  & $1.53 \pm 0.73$ \\
Hf$_4$--O       & $27.06 \pm 2.02$ & $6.36 \pm 1.51$  & $6.15 \pm 1.37$ \\
Hf$_3$Si--O     & $0.00 \pm 0.00$  & $2.04 \pm 0.43$  & $1.84 \pm 0.59$ \\
Other           & $0.32 \pm 0.19$  & $0.02 \pm 0.04$  & $0.03 \pm 0.04$ \\
\end{tabular}
\end{ruledtabular}
\end{table*}

\noindent total PDF is dominated  by Hf--Hf correlations, with a maximum near 3.4~\AA\ and a shoulder slightly below 4 \AA.

The \sdh\ models contain an additional strong Si--O peak near 1.62~\AA. This peak corresponds to SiO$_4$ tetrahedra. However, because Si and O have much smaller x-ray scattering weights than Hf, the Si--O correlation does not dominate the measured total GIPDF and appears as a small prepeak before the Hf--O peak. These partial PDFs therefore show why the total GIPDF alone cannot uniquely identify all local structural motifs. The measured PDF is dominated by Hf--O and Hf--Hf correlations, whereas the main structural role of \silica\ is seen more clearly in the model-derived Si--O, Si--Si, and Hf--Si partial correlations.

The mixed-cation partials are especially useful for distinguishing two limiting structural pictures for \sdh. In one limit, SiO$_4$ tetrahedra could cluster into \silica-rich regions that are only weakly connected to the Hf--O network. In the other limit, the SiO$_4$ units could be dispersed through the film and linked primarily to Hf-centered polyhedra through Si--O--Hf bridges. The presence of both Hf--Si and Si--Si correlations in Fig.~\ref{fig:partial_pdfs_hfo2_sdh} indicates that the experimental GIPDF-constrained models contain a mixed network rather than a purely \hafnia-like network plus isolated \silica\ units. A more direct test of this picture is oxygen-bridge analysis i.e. analyzing the cation identities around each oxygen atom, because the fractions of Hf--O--Hf, Hf--O--Si, and Si--O--Si bridges measure how \silica\ is incorporated into the \hafnia\ network.

The oxygen-bridge analysis, summarized in Table~\ref{tab:oxygen_bridge}, supports the dispersed mixed-network picture. In pure \hafnia, oxygen atoms are dominated by Hf-rich high-connectivity environments: most O atoms are bonded to three Hf atoms, and a substantial fraction are bonded to four Hf atoms. In contrast, the \sdh\ models contain large populations of mixed Hf--O--Si and Hf$_2$Si--O environments, while purely Si--O--Si bridges remain comparatively small. Thus, the SiO$_4$ units do not appear primarily as isolated \silica-rich clusters; rather, they are incorporated into the Hf--O network through mixed Si--O--Hf linkages. This incorporation lowers the fraction of highly connected Hf-rich oxygen environments, especially Hf$_4$--O, and replaces them with lower-connectivity mixed bridges. The main structural role of \silica\ is therefore to open and chemically mix the amorphous \hafnia\ network, rather than to form a separate \silica-like phase. The similarity of the 150$^\circ$C and 400$^\circ$C \sdh\ bridge populations further indicates that this mixed topology is established by composition and is only weakly modified by annealing.

\subsection{Coordination statistics and polyhedral connectivity}

\begin{figure}[t]
	\centering
	\includegraphics[width=\linewidth]{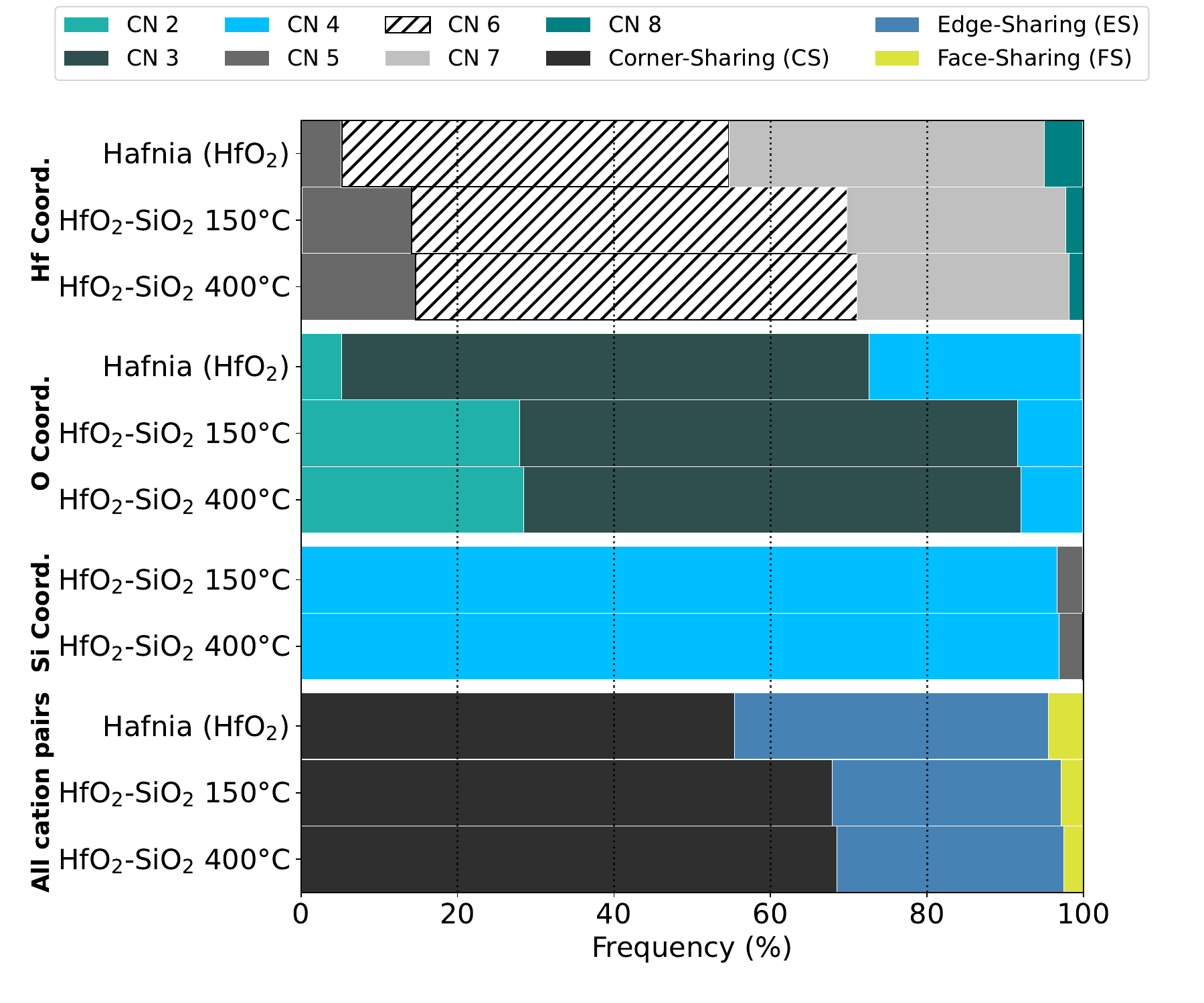}
	\caption{Coordination and polyhedral-connectivity descriptors from the models. Pure \hafnia\ contains mostly sixfold and sevenfold Hf, together with many edge-sharing and face-sharing Hf-centered polyhedra. In \sdh, Si remains nearly fourfold coordinated and the total cation--cation network shifts toward corner-sharing connections. The 400$^\circ$C annealed \sdh\ model shows only small changes relative to the 150$^\circ$C sample.}
	\label{fig:coord_connectivity_hfo2_sdh}
\end{figure}

Fig.~\ref{fig:coord_connectivity_hfo2_sdh} shows the coordination distribution around Hf, O and Si atoms computed from the models. For pure \hafnia, the average Hf coordination is 6.45, with most Hf atoms appearing in sixfold and sevenfold oxygen coordination. The distribution is broad: approximately 49\% of Hf atoms are sixfold coordinated and 40\% are sevenfold coordinated, with smaller populations of fivefold and eightfold Hf. Oxygen atoms are also highly coordinated, with an average cation coordination of 3.23. Most O atoms are threefold coordinated, and a substantial fraction are fourfold coordinated. This confirms that amorphous \hafnia\ is a dense, high-coordination oxide network rather than a tetrahedral glass network.

The 27\% \sdh\ models show a clear change in local structure. The average Hf coordination decreases from 6.45 in pure \hafnia\ to 6.18 and 6.16 in the 150$^\circ$C and 400$^\circ$C \sdh\ films, respectively. The population of sevenfold and eightfold Hf decreases, while the population of fivefold Hf increases. At the same time, Si remains almost entirely fourfold coordinated: approximately 97\% of Si atoms form SiO$_4$ units in both \sdh\ samples. Thus, \silica\ incorporation does not simply dilute the Hf network; it introduces a different local building block into the amorphous structure.

The oxygen coordination changes even more strongly. The average O cation-coordination decreases from 3.23 in pure \hafnia\ to about 2.80 in \sdh. In pure \hafnia, only about 5\% of O atoms are twofold coordinated, whereas in \sdh\ nearly 28\% of O atoms are twofold coordinated. Conversely, the fraction of fourfold-coordinated O decreases from about 27\% in pure \hafnia\ to about 8\% in \sdh. This indicates that \silica\ doping opens the network by replacing some highly connected Hf--O--Hf environments with lower-connectivity Si--O--Hf and Si--O--Si environments. This change in oxygen connectivity is central to the structural role of \silica: a twofold oxygen can act as a simple bridge between neighboring cation polyhedra, while threefold and fourfold oxygen atoms support a more compact, highly connected Hf-rich network.

Annealing from 150$^\circ$C to 400$^\circ$C produces only small changes in coordination. The Hf and Si coordinations are essentially unchanged within the model-to-model variation. The O coordination also changes only weakly, with a small increase in twofold O and a small decrease in fourfold O. Therefore, the dominant coordination change is caused by \silica\ addition, while annealing mainly modifies the topology of connections between already-formed local units.

The connectivity between neighboring cation-centered polyhedra is also shown in Fig. \ref{fig:coord_connectivity_hfo2_sdh}. We classify two neighboring cation-centered polyhedra as corner-sharing (CS), edge-sharing (ES), or face-sharing (FS) when they share one, two, or three oxygen atoms, respectively. In pure \hafnia, the Hf-centered polyhedra form a compact network: about 55\% of connected Hf--Hf pairs are corner-sharing, about 40\% are edge-sharing, and about 4.5\% are face-sharing. The large population of ES and FS connections reflects the dense packing of high-coordination Hf polyhedra. Similar connections between Hf coordination, Hf--Hf correlations, and polyhedral packing have been noted in diffraction and simulation studies of amorphous \hafnia, where HfO$_6$ and HfO$_7$ units form both corner- and edge-sharing motifs depending on density and preparation route \cite{gallington2017structure,nguyen2019molecular}.

The \sdh\ films contain a substantially larger fraction of corner-sharing connections when all cation--cation pairs are considered. In the 150$^\circ$C \sdh\ film, about 68\% of all connected cation pairs are corner-sharing, about 29\% are edge-sharing, and about 2.8\% are face-sharing. The 400$^\circ$C film is similar, but with a small further shift toward corner-sharing connections. This change is not caused by a large change in Hf--Hf connectivity alone. In fact, the Hf--Hf connections in \sdh\ still contain substantial ES and FS populations. Instead, the change comes mainly from the mixed and Si-containing connections: Hf--Si pairs are mostly corner-sharing, and Si--Si pairs are essentially entirely corner-sharing. Thus, \silica\ incorporation reroutes part of the network through SiO$_4$ tetrahedra, reducing the total fraction of compact ES/FS motifs.

This result gives a clear structural picture of how \silica\ stabilizes the amorphous \hafnia-based network. Pure \hafnia\ favors a dense arrangement of high-coordination Hf polyhedra with many shared edges and faces. Adding \silica\ introduces tetrahedral units that connect primarily through corners. This lowers the average oxygen coordination, reduces the fraction of compact shared-polyhedron motifs, and increases the openness of the cation--oxygen network. The annealed \sdh\ sample preserves this mixed-network topology, with only a small reduction in the FS population. This behavior is consistent with the experimentally observed amorphous stability of 27\% \sdh\ coatings after heat treatment \cite{Craig2019PRL}.

\subsection{Bond-angle distributions}
\begin{figure}[h]
	\centering
	\includegraphics[width=\linewidth]{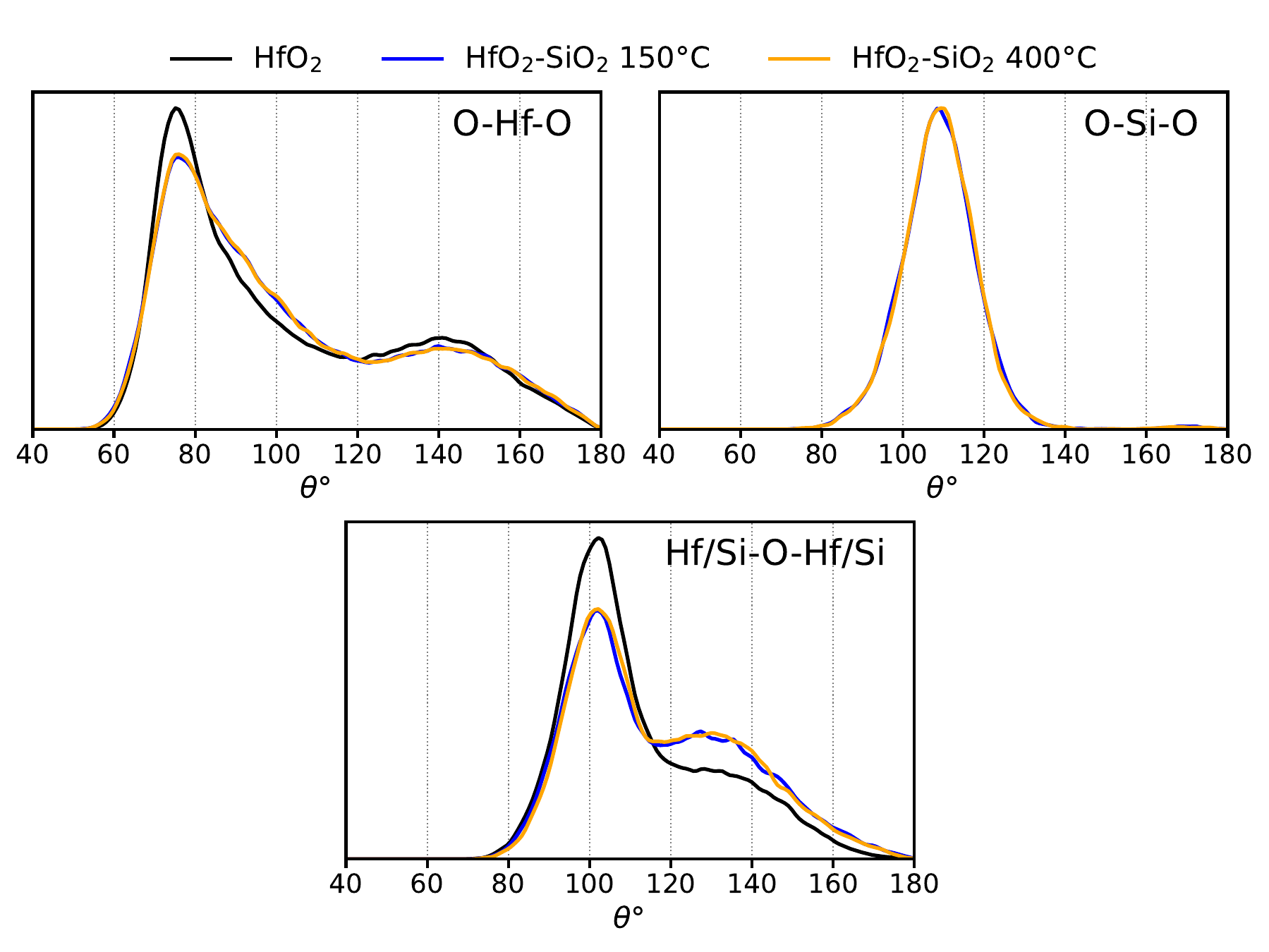}
	\caption{Bond-angle distributions from the atomic models. The O--Hf--O distributions are broad, as expected for high-coordination Hf-centered polyhedra. The O--Si--O distributions in \sdh\ are centered near the tetrahedral angle, confirming that Si mainly forms SiO$_4$ units. The cation--O--cation distributions shift to larger average angles in \sdh, consistent with a more open mixed network.}
	\label{fig:angles_hfo2_sdh}
\end{figure}

The bond-angle distributions provide a second view of the local topology. Figure~\ref{fig:angles_hfo2_sdh} shows the O--Hf--O, O--Si--O, and cation--O--cation angle distributions. The O--Hf--O distributions are broad in all samples, with maxima near 75$^\circ$ and average angles near 102$^\circ$. This broad distribution is expected for high-coordination Hf-centered polyhedra in an amorphous oxide. The similarity of the O--Hf--O distributions in pure \hafnia\ and \sdh\ shows that \silica\ doping does not fundamentally change the internal geometry of individual Hf-centered polyhedra.

In contrast, the O--Si--O distribution in \sdh\ is much narrower and centered near the tetrahedral angle. The average O--Si--O angle is about 108.6$^\circ$ in both \sdh\ samples. This confirms that Si enters the structure primarily as SiO$_4$ tetrahedra. The persistence of this distribution after 400$^\circ$C annealing shows that these tetrahedral units are stable during the heat treatment considered here.

The cation--O--cation angle distribution is broader and more sensitive to the mixed-network topology. The average cation--O--cation angle increases from about 112$^\circ$ in pure \hafnia\ to about 117$^\circ$ in \sdh. This increase is consistent with the formation of a more open network when SiO$_4$ tetrahedra are incorporated into the Hf--O framework. The difference between the 150$^\circ$C and 400$^\circ$C \sdh\ films is again small, indicating that annealing changes the medium-range packing more than the local bond geometry.

\subsection{Intermediate-range structure and structural picture}

The GIPDFs and structural models also reveal differences beyond the first two coordination shells. The measured and modeled PDFs contain broad features near 6~\AA\ and 8--9~\AA, indicating intermediate-range correlations in the amorphous network. These features arise primarily from correlations between Hf-centered polyhedra and from the packing and connectivity of the constituent structural units. The pure-\hafnia\ model consists of a dense network of interconnected HfO$_n$ polyhedra, whereas the \sdh\ models contain a mixed network in which SiO$_4$ tetrahedra interrupt and redirect the Hf--O framework.

\begin{figure}[t]
	\centering
	\includegraphics[width=\linewidth]{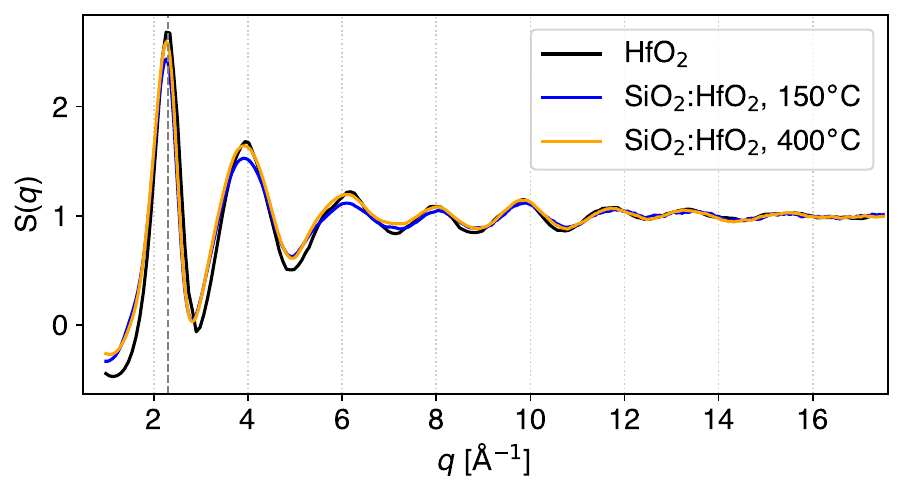}
	\caption{Measured structure factors $S(q)$ from thin films of \hafnia\ and \sdh. The dashed grey line marks the first sharp diffraction peak (FSDP) near 2.3~\AA$^{-1}$. The same data-processing protocol was used to obtain all curves. The observed differences are robust to small changes in the data-processing choices.}
	\label{fig:sq_fsdp}
\end{figure}

These intermediate-range structural differences are also evident in the measured structure factors shown in Fig.~\ref{fig:sq_fsdp}. All three films exhibit a first sharp diffraction peak (FSDP) near $q=2.2$--$2.3$~\AA$^{-1}$, consistent with medium-range correlations in the amorphous network. FSDP features in amorphous materials are commonly associated with correlations extending beyond the nearest-neighbor coordination shell and with network topology rather than with a single interatomic distance \cite{elliott1991medium,shyam2016measurement,mishkin2023hidden}. Gaussian fits give FSDP positions of 2.285~\AA$^{-1}$ for pure \hafnia, 2.225~\AA$^{-1}$ for \sdh\ at 150$^\circ$C, and 2.230~\AA$^{-1}$ for \sdh\ at 400$^\circ$C. Thus, incorporation of 27\% \silica\ produces a modest shift of the FSDP toward lower $q$, whereas subsequent annealing produces essentially no further shift in its position.

The FSDP line shape changes more substantially upon \silica\ incorporation. The fitted FWHM increases from 0.385~\AA$^{-1}$ in pure \hafnia\ to 0.498 and 0.499~\AA$^{-1}$ in the 150$^\circ$C and 400$^\circ$C \sdh\ films, respectively. The narrower FSDP of pure \hafnia\ therefore indicates that the intermediate-range correlations responsible for this feature remain coherent over a greater spatial extent. Using $2\pi/\Delta q$ as an approximate measure of this coherence length gives values of approximately 16.3~\AA\ for pure \hafnia\ and 12.6~\AA\ for both \sdh\ films. At the same time, the fitted FSDP amplitude increases from 1.99 in pure \hafnia\ to 2.22 and 2.39 in the 150$^\circ$C and 400$^\circ$C \sdh\ films, respectively. Because the total x-ray structure factor is compositionally weighted, the larger FSDP amplitude in \sdh\ should not by itself be interpreted as greater medium-range order than in pure \hafnia; rather, it indicates a greater prominence of the x-ray-weighted correlations contributing to the FSDP. The simultaneous increase in peak amplitude and broadening upon \silica\ addition therefore suggests a reorganization of the intermediate-range network rather than a simple increase or decrease in structural order.

In contrast, the two \sdh\ films have nearly identical FSDP positions and widths, showing that annealing from 150$^\circ$C to 400$^\circ$C does not substantially alter the characteristic length scale or spatial extent of their intermediate-range correlations. The modest increase in FSDP amplitude upon annealing, from 2.22 to 2.39, instead suggests a strengthening or increased definition of correlations within an otherwise similar mixed Hf--O--Si network. Together with the similarities in the higher-$q$ oscillations, these results indicate that the dominant intermediate-range structural reorganization is produced by \silica\ incorporation, while the subsequent annealing step causes a comparatively subtle relaxation of the mixed network.

\subsection{Implications for mechanical loss}
\label{sec:structure_loss}

The structural trends observed here also suggest a useful way to think about the temperature dependence of mechanical loss in amorphous oxide coatings. The main effect of \silica\ incorporation is not a large change in the Hf--O bond length or in the internal geometry of individual Hf-centered polyhedra. Instead, \silica\ changes the topology of the amorphous network. It lowers the average Hf and O coordination, introduces stable SiO$_4$ tetrahedra, increases the fraction of twofold oxygen atoms, and converts part of the dense Hf--O--Hf network into mixed Hf--O--Si and Si--O--Si linkages. At the same time, the total cation network shifts away from compact edge- and face-sharing motifs and toward more corner-sharing connectivity. Annealing from 150$^\circ$C to 400$^\circ$C produces a smaller version of the same topological change: the local coordination statistics and bond-angle distributions remain nearly unchanged, while the compact shared-polyhedron population decreases slightly and the corner-sharing population increases. Thus, both \silica\ addition and annealing can be viewed primarily as network-topology changes rather than changes in nearest-neighbor bond geometry.

Across several amorphous oxides, similar topology changes appear to correlate with room-temperature mechanical loss. In amorphous \tantala-based coatings, atomistic modeling and total-scattering measurements have connected high mechanical loss with compact polyhedral environments, especially edge- and face-sharing motifs, while annealing reduces these motifs and lowers room-temperature loss \cite{prasai2019high,prasai2021annealing}. Atomistic simulations of amorphous \tantala\ provide a consistent microscopic picture: Puosi \textit{et al.} found that polyhedra containing atoms with the largest irreversible rearrangements are enriched in edge- and face-sharing connections and depleted in corner-sharing connections \cite{puosi2020nonlocal}. In this view, compact shared-polyhedron motifs provide structurally frustrated local environments that can support the thermally activated rearrangements responsible for room-temperature dissipation. The reduction of edge- and face-sharing motifs in \sdh\ is therefore structurally favorable from the standpoint of room-temperature loss, even though direct room-temperature loss measurements on the exact films studied here are not available.

The same structural descriptors, however, need not have the same effect at cryogenic temperatures. \silica\ provides the clearest example: its network is dominated by corner-sharing SiO$_4$ tetrahedra and has exceptionally low room-temperature loss, but it exhibits a pronounced cryogenic loss peak. Conversely, 27\% \sdh\ has substantially lower cryogenic loss than \silica\ below approximately 40~K, even though the Hf-containing part of the network still contains a significant population of compact shared-polyhedron motifs \cite{Craig2019PRL}. This contrast suggests that the structural motifs that suppress room-temperature dissipation are not necessarily the same motifs that suppress cryogenic dissipation. A plausible interpretation is that edge- and face-sharing environments are associated with higher-barrier rearrangements that contribute strongly at room temperature, whereas corner-sharing networks can support lower-barrier angular or rotational rearrangements that become important at cryogenic temperatures.

This distinction is also consistent with atomistic studies of low-temperature rearrangements. Damart and Rodney found that two-level systems in amorphous \silica\ involve coordinated displacements of atoms across connected SiO$_4$ tetrahedra, often without bond breaking \cite{damart2018atomistic}. Puosi \textit{et al.} similarly reported an anomalous contribution from oxygen motion near the low-temperature loss peak in amorphous \tantala\ \cite{puosi2020nonlocal}. These results suggest that cryogenic loss may be weighted toward low-barrier, oxygen-mediated rearrangements within corner-connected polyhedral networks, while room-temperature loss may be more strongly associated with compact edge- and face-sharing environments. In this sense, annealing and compositional modification may redistribute the spectrum of relaxation pathways rather than simply remove all mechanically active defects.

Within this broader picture, the present \hafnia-based models provide an important structural reference point. Pure amorphous \hafnia\ is a dense, high-coordination network with many edge- and face-sharing Hf-centered polyhedra. Adding 27\% \silica\ opens this network by introducing SiO$_4$ tetrahedra and mixed Si--O--Hf bridges, reducing the population of highly connected Hf-rich oxygen environments and increasing corner-sharing connectivity. This topology is consistent with improved amorphous stability and with the low cryogenic loss previously reported for \sdh\ coatings \cite{Craig2019PRL}. At the same time, the remaining edge- and face-sharing Hf-centered motifs may still be relevant for room-temperature dissipation. The comparison therefore points to a materials-design principle for amorphous optical coatings: reducing room-temperature loss may require suppressing compact edge- and face-sharing motifs, while reducing cryogenic loss may require controlling the flexibility and oxygen-mediated rearrangements of the corner-connected network.

\section{CONCLUSIONS}\label{sec:conclusions}

We have combined GIPDF measurements with experimentally constrained atomic modeling to determine the short- and intermediate-range structure of amorphous \hafnia\ and 27\% \sdh\ optical thin film coatings. The model ensembles reproduce the main measured GIPDF features and provide chemically reasonable structures from which partial PDFs, coordination statistics, oxygen-bridge populations, bond-angle distributions, and polyhedral-connectivity descriptors were extracted. 

The main structural effect of \silica\ incorporation is to change the topology of the amorphous \hafnia\ network. Pure amorphous \hafnia\ forms a dense, high-coordination network dominated by sixfold and sevenfold Hf atoms, threefold and fourfold oxygen atoms, and substantial edge- and face-sharing connectivity between Hf-centered polyhedra. In contrast, 27\% \sdh\ contains nearly fourfold-coordinated Si in SiO$_4$ tetrahedra, lower average Hf and O coordination, and a larger fraction of twofold oxygen atoms. Oxygen-bridge analysis shows that \silica\ reduces highly connected Hf-rich environments, especially Hf$_4$--O, and replaces them with mixed Hf--O--Si and Hf$_2$Si--O environments. Thus, the doped films are best described as chemically mixed amorphous networks in which SiO$_4$ units are incorporated into the Hf--O framework, rather than as isolated \silica-rich regions embedded in a \hafnia-like matrix.

Annealing the 27\% \sdh\ film from 150$^\circ$C to 400$^\circ$C produces much smaller structural changes than adding \silica. The Hf and Si coordination distributions, oxygen-bridge populations, bond-angle distributions, and main pair correlations remain nearly unchanged within the model-to-model variation, while the polyhedral connectivity shows only a small shift away from compact shared-polyhedron motifs and toward corner sharing. These results support a structural picture in which \silica\ stabilizes amorphous \hafnia-based coatings by opening and chemically mixing the network. They also provide a concrete atomic-scale basis for interpreting mechanical loss: compact edge- and face-sharing motifs may be relevant to room-temperature dissipation, while mixed corner-connected and oxygen-mediated environments may influence the lower-barrier relaxation pathways active at cryogenic temperatures.

\section{DATA AVAILABILITY}
Data are available from the corresponding author upon reasonable request.

\section{ACKNOWLEDGMENTS}
KP acknowledges support from the National Science Foundation (NSF), grant number 2513491, and the Gordon and Betty Moore Foundation (GBMF), grant GBMF6793.02. MMF acknowledges support from the NSF under grant numbers 2309289, 2309086, and 2513483, and from the GBMF under grant GBMF6793.02. KHL acknowledges support from the National Research Foundation of Korea (NRF), funded by the Korean government (MSIT), under grant No. RS-2024-00455482. This paper has LIGO document No. P2600432.

\bibliography{prasai-refs-M18}

\begin{thebibliography}{57}%
\makeatletter
\providecommand \@ifxundefined [1]{%
 \@ifx{#1\undefined}
}%
\providecommand \@ifnum [1]{%
 \ifnum #1\expandafter \@firstoftwo
 \else \expandafter \@secondoftwo
 \fi
}%
\providecommand \@ifx [1]{%
 \ifx #1\expandafter \@firstoftwo
 \else \expandafter \@secondoftwo
 \fi
}%
\providecommand \natexlab [1]{#1}%
\providecommand \enquote  [1]{``#1''}%
\providecommand \bibnamefont  [1]{#1}%
\providecommand \bibfnamefont [1]{#1}%
\providecommand \citenamefont [1]{#1}%
\providecommand \href@noop [0]{\@secondoftwo}%
\providecommand \href [0]{\begingroup \@sanitize@url \@href}%
\providecommand \@href[1]{\@@startlink{#1}\@@href}%
\providecommand \@@href[1]{\endgroup#1\@@endlink}%
\providecommand \@sanitize@url [0]{\catcode `\\12\catcode `\$12\catcode `\&12\catcode `\#12\catcode `\^12\catcode `\_12\catcode `\%12\relax}%
\providecommand \@@startlink[1]{}%
\providecommand \@@endlink[0]{}%
\providecommand \url  [0]{\begingroup\@sanitize@url \@url }%
\providecommand \@url [1]{\endgroup\@href {#1}{\urlprefix }}%
\providecommand \urlprefix  [0]{URL }%
\providecommand \Eprint [0]{\href }%
\providecommand \doibase [0]{https://doi.org/}%
\providecommand \selectlanguage [0]{\@gobble}%
\providecommand \bibinfo  [0]{\@secondoftwo}%
\providecommand \bibfield  [0]{\@secondoftwo}%
\providecommand \translation [1]{[#1]}%
\providecommand \BibitemOpen [0]{}%
\providecommand \bibitemStop [0]{}%
\providecommand \bibitemNoStop [0]{.\EOS\space}%
\providecommand \EOS [0]{\spacefactor3000\relax}%
\providecommand \BibitemShut  [1]{\csname bibitem#1\endcsname}%
\let\auto@bib@innerbib\@empty
\bibitem [{\citenamefont {Martynov}\ \emph {et~al.}(2016)\citenamefont {Martynov}, \citenamefont {Hall}, \citenamefont {Abbott}, \citenamefont {Abbott}, \citenamefont {Abbott}, \citenamefont {Adams}, \citenamefont {Adhikari}, \citenamefont {Anderson}, \citenamefont {Anderson}, \citenamefont {Arai} \emph {et~al.}}]{martynov2016sensitivity}%
  \BibitemOpen
  \bibfield  {author} {\bibinfo {author} {\bibfnamefont {D.~V.}\ \bibnamefont {Martynov}}, \bibinfo {author} {\bibfnamefont {E.}~\bibnamefont {Hall}}, \bibinfo {author} {\bibfnamefont {B.}~\bibnamefont {Abbott}}, \bibinfo {author} {\bibfnamefont {R.}~\bibnamefont {Abbott}}, \bibinfo {author} {\bibfnamefont {T.}~\bibnamefont {Abbott}}, \bibinfo {author} {\bibfnamefont {C.}~\bibnamefont {Adams}}, \bibinfo {author} {\bibfnamefont {R.}~\bibnamefont {Adhikari}}, \bibinfo {author} {\bibfnamefont {R.}~\bibnamefont {Anderson}}, \bibinfo {author} {\bibfnamefont {S.}~\bibnamefont {Anderson}}, \bibinfo {author} {\bibfnamefont {K.}~\bibnamefont {Arai}}, \emph {et~al.},\ }\href@noop {} {\bibfield  {journal} {\bibinfo  {journal} {Physical Review D}\ }\textbf {\bibinfo {volume} {93}},\ \bibinfo {pages} {112004} (\bibinfo {year} {2016})}\BibitemShut {NoStop}%
\bibitem [{\citenamefont {Buikema}\ \emph {et~al.}(2020)\citenamefont {Buikema}, \citenamefont {Cahillane}, \citenamefont {Mansell}, \citenamefont {Blair}, \citenamefont {Abbott}, \citenamefont {Adams}, \citenamefont {Adhikari}, \citenamefont {Ananyeva}, \citenamefont {Appert}, \citenamefont {Arai} \emph {et~al.}}]{buikema2020sensitivity}%
  \BibitemOpen
  \bibfield  {author} {\bibinfo {author} {\bibfnamefont {A.}~\bibnamefont {Buikema}}, \bibinfo {author} {\bibfnamefont {C.}~\bibnamefont {Cahillane}}, \bibinfo {author} {\bibfnamefont {G.}~\bibnamefont {Mansell}}, \bibinfo {author} {\bibfnamefont {C.}~\bibnamefont {Blair}}, \bibinfo {author} {\bibfnamefont {R.}~\bibnamefont {Abbott}}, \bibinfo {author} {\bibfnamefont {C.}~\bibnamefont {Adams}}, \bibinfo {author} {\bibfnamefont {R.}~\bibnamefont {Adhikari}}, \bibinfo {author} {\bibfnamefont {A.}~\bibnamefont {Ananyeva}}, \bibinfo {author} {\bibfnamefont {S.}~\bibnamefont {Appert}}, \bibinfo {author} {\bibfnamefont {K.}~\bibnamefont {Arai}}, \emph {et~al.},\ }\href@noop {} {\bibfield  {journal} {\bibinfo  {journal} {Physical Review D}\ }\textbf {\bibinfo {volume} {102}},\ \bibinfo {pages} {062003} (\bibinfo {year} {2020})}\BibitemShut {NoStop}%
\bibitem [{\citenamefont {Harry}\ \emph {et~al.}(2006)\citenamefont {Harry}, \citenamefont {Abernathy}, \citenamefont {Becerra-Toledo}, \citenamefont {Armandula}, \citenamefont {Black}, \citenamefont {Dooley}, \citenamefont {Eichenfield}, \citenamefont {Nwabugwu}, \citenamefont {Villar}, \citenamefont {Crooks} \emph {et~al.}}]{harry2006titania}%
  \BibitemOpen
  \bibfield  {author} {\bibinfo {author} {\bibfnamefont {G.~M.}\ \bibnamefont {Harry}}, \bibinfo {author} {\bibfnamefont {M.~R.}\ \bibnamefont {Abernathy}}, \bibinfo {author} {\bibfnamefont {A.~E.}\ \bibnamefont {Becerra-Toledo}}, \bibinfo {author} {\bibfnamefont {H.}~\bibnamefont {Armandula}}, \bibinfo {author} {\bibfnamefont {E.}~\bibnamefont {Black}}, \bibinfo {author} {\bibfnamefont {K.}~\bibnamefont {Dooley}}, \bibinfo {author} {\bibfnamefont {M.}~\bibnamefont {Eichenfield}}, \bibinfo {author} {\bibfnamefont {C.}~\bibnamefont {Nwabugwu}}, \bibinfo {author} {\bibfnamefont {A.}~\bibnamefont {Villar}}, \bibinfo {author} {\bibfnamefont {D.}~\bibnamefont {Crooks}}, \emph {et~al.},\ }\href@noop {} {\bibfield  {journal} {\bibinfo  {journal} {Classical and Quantum Gravity}\ }\textbf {\bibinfo {volume} {24}},\ \bibinfo {pages} {405} (\bibinfo {year} {2006})}\BibitemShut {NoStop}%
\bibitem [{\citenamefont {Steinlechner}(2018)}]{steinlechner2018development}%
  \BibitemOpen
  \bibfield  {author} {\bibinfo {author} {\bibfnamefont {J.}~\bibnamefont {Steinlechner}},\ }\href@noop {} {\bibfield  {journal} {\bibinfo  {journal} {Philosophical Transactions of the Royal Society A: Mathematical, Physical and Engineering Sciences}\ }\textbf {\bibinfo {volume} {376}},\ \bibinfo {pages} {20170282} (\bibinfo {year} {2018})}\BibitemShut {NoStop}%
\bibitem [{\citenamefont {Amato}\ \emph {et~al.}(2019)\citenamefont {Amato}, \citenamefont {Terreni}, \citenamefont {Dolique}, \citenamefont {Forest}, \citenamefont {Gemme}, \citenamefont {Granata}, \citenamefont {Mereni}, \citenamefont {Michel}, \citenamefont {Pinard}, \citenamefont {Sassolas} \emph {et~al.}}]{amato2019optical}%
  \BibitemOpen
  \bibfield  {author} {\bibinfo {author} {\bibfnamefont {A.}~\bibnamefont {Amato}}, \bibinfo {author} {\bibfnamefont {S.}~\bibnamefont {Terreni}}, \bibinfo {author} {\bibfnamefont {V.}~\bibnamefont {Dolique}}, \bibinfo {author} {\bibfnamefont {D.}~\bibnamefont {Forest}}, \bibinfo {author} {\bibfnamefont {G.}~\bibnamefont {Gemme}}, \bibinfo {author} {\bibfnamefont {M.}~\bibnamefont {Granata}}, \bibinfo {author} {\bibfnamefont {L.}~\bibnamefont {Mereni}}, \bibinfo {author} {\bibfnamefont {C.}~\bibnamefont {Michel}}, \bibinfo {author} {\bibfnamefont {L.}~\bibnamefont {Pinard}}, \bibinfo {author} {\bibfnamefont {B.}~\bibnamefont {Sassolas}}, \emph {et~al.},\ }\href@noop {} {\bibfield  {journal} {\bibinfo  {journal} {Journal of Physics: Materials}\ }\textbf {\bibinfo {volume} {2}},\ \bibinfo {pages} {035004} (\bibinfo {year} {2019})}\BibitemShut {NoStop}%
\bibitem [{\citenamefont {Granata}\ \emph {et~al.}(2020)\citenamefont {Granata}, \citenamefont {Amato}, \citenamefont {Balzarini}, \citenamefont {Canepa}, \citenamefont {Degallaix}, \citenamefont {Forest}, \citenamefont {Dolique}, \citenamefont {Mereni}, \citenamefont {Michel}, \citenamefont {Pinard} \emph {et~al.}}]{granata2020amorphous}%
  \BibitemOpen
  \bibfield  {author} {\bibinfo {author} {\bibfnamefont {M.}~\bibnamefont {Granata}}, \bibinfo {author} {\bibfnamefont {A.}~\bibnamefont {Amato}}, \bibinfo {author} {\bibfnamefont {L.}~\bibnamefont {Balzarini}}, \bibinfo {author} {\bibfnamefont {M.}~\bibnamefont {Canepa}}, \bibinfo {author} {\bibfnamefont {J.}~\bibnamefont {Degallaix}}, \bibinfo {author} {\bibfnamefont {D.}~\bibnamefont {Forest}}, \bibinfo {author} {\bibfnamefont {V.}~\bibnamefont {Dolique}}, \bibinfo {author} {\bibfnamefont {L.}~\bibnamefont {Mereni}}, \bibinfo {author} {\bibfnamefont {C.}~\bibnamefont {Michel}}, \bibinfo {author} {\bibfnamefont {L.}~\bibnamefont {Pinard}}, \emph {et~al.},\ }\href@noop {} {\bibfield  {journal} {\bibinfo  {journal} {Classical and Quantum Gravity}\ }\textbf {\bibinfo {volume} {37}},\ \bibinfo {pages} {095004} (\bibinfo {year} {2020})}\BibitemShut {NoStop}%
\bibitem [{\citenamefont {Miller}\ \emph {et~al.}(2015)\citenamefont {Miller}, \citenamefont {Barsotti}, \citenamefont {Vitale}, \citenamefont {Fritschel}, \citenamefont {Evans},\ and\ \citenamefont {Sigg}}]{miller2015prospects}%
  \BibitemOpen
  \bibfield  {author} {\bibinfo {author} {\bibfnamefont {J.}~\bibnamefont {Miller}}, \bibinfo {author} {\bibfnamefont {L.}~\bibnamefont {Barsotti}}, \bibinfo {author} {\bibfnamefont {S.}~\bibnamefont {Vitale}}, \bibinfo {author} {\bibfnamefont {P.}~\bibnamefont {Fritschel}}, \bibinfo {author} {\bibfnamefont {M.}~\bibnamefont {Evans}},\ and\ \bibinfo {author} {\bibfnamefont {D.}~\bibnamefont {Sigg}},\ }\href@noop {} {\bibfield  {journal} {\bibinfo  {journal} {Physical Review D}\ }\textbf {\bibinfo {volume} {91}},\ \bibinfo {pages} {062005} (\bibinfo {year} {2015})}\BibitemShut {NoStop}%
\bibitem [{\citenamefont {{The LIGO Scientific Collaboration}}(2020)}]{AplusWP}%
  \BibitemOpen
  \bibfield  {author} {\bibinfo {author} {\bibnamefont {{The LIGO Scientific Collaboration}}},\ }\href {https://dcc.ligo.org/LIGO-T2000407/public} {\bibinfo {title} {Instrument science white paper 2020}},\ \bibinfo {howpublished} {LIGO document, Report No. T2000407} (\bibinfo {year} {2020})\BibitemShut {NoStop}%
\bibitem [{\citenamefont {Punturo}\ \emph {et~al.}(2010)\citenamefont {Punturo}, \citenamefont {Abernathy}, \citenamefont {Acernese}, \citenamefont {Allen}, \citenamefont {Andersson}, \citenamefont {Arun}, \citenamefont {Barone}, \citenamefont {Barr}, \citenamefont {Barsuglia}, \citenamefont {Beker} \emph {et~al.}}]{punturo2010einstein}%
  \BibitemOpen
  \bibfield  {author} {\bibinfo {author} {\bibfnamefont {M.}~\bibnamefont {Punturo}}, \bibinfo {author} {\bibfnamefont {M.}~\bibnamefont {Abernathy}}, \bibinfo {author} {\bibfnamefont {F.}~\bibnamefont {Acernese}}, \bibinfo {author} {\bibfnamefont {B.}~\bibnamefont {Allen}}, \bibinfo {author} {\bibfnamefont {N.}~\bibnamefont {Andersson}}, \bibinfo {author} {\bibfnamefont {K.}~\bibnamefont {Arun}}, \bibinfo {author} {\bibfnamefont {F.}~\bibnamefont {Barone}}, \bibinfo {author} {\bibfnamefont {B.}~\bibnamefont {Barr}}, \bibinfo {author} {\bibfnamefont {M.}~\bibnamefont {Barsuglia}}, \bibinfo {author} {\bibfnamefont {M.}~\bibnamefont {Beker}}, \emph {et~al.},\ }\href@noop {} {\bibfield  {journal} {\bibinfo  {journal} {Classical and Quantum Gravity}\ }\textbf {\bibinfo {volume} {27}},\ \bibinfo {pages} {194002} (\bibinfo {year} {2010})}\BibitemShut {NoStop}%
\bibitem [{\citenamefont {Hild}(2012)}]{hild2012beyond}%
  \BibitemOpen
  \bibfield  {author} {\bibinfo {author} {\bibfnamefont {S.}~\bibnamefont {Hild}},\ }\href@noop {} {\bibfield  {journal} {\bibinfo  {journal} {Classical and Quantum Gravity}\ }\textbf {\bibinfo {volume} {29}},\ \bibinfo {pages} {124006} (\bibinfo {year} {2012})}\BibitemShut {NoStop}%
\bibitem [{\citenamefont {Hall}\ \emph {et~al.}(2021)\citenamefont {Hall}, \citenamefont {Kuns}, \citenamefont {Smith}, \citenamefont {Bai}, \citenamefont {Wipf}, \citenamefont {Biscans}, \citenamefont {Adhikari}, \citenamefont {Arai}, \citenamefont {Ballmer}, \citenamefont {Barsotti} \emph {et~al.}}]{hall2021gravitational}%
  \BibitemOpen
  \bibfield  {author} {\bibinfo {author} {\bibfnamefont {E.~D.}\ \bibnamefont {Hall}}, \bibinfo {author} {\bibfnamefont {K.}~\bibnamefont {Kuns}}, \bibinfo {author} {\bibfnamefont {J.~R.}\ \bibnamefont {Smith}}, \bibinfo {author} {\bibfnamefont {Y.}~\bibnamefont {Bai}}, \bibinfo {author} {\bibfnamefont {C.}~\bibnamefont {Wipf}}, \bibinfo {author} {\bibfnamefont {S.}~\bibnamefont {Biscans}}, \bibinfo {author} {\bibfnamefont {R.~X.}\ \bibnamefont {Adhikari}}, \bibinfo {author} {\bibfnamefont {K.}~\bibnamefont {Arai}}, \bibinfo {author} {\bibfnamefont {S.}~\bibnamefont {Ballmer}}, \bibinfo {author} {\bibfnamefont {L.}~\bibnamefont {Barsotti}}, \emph {et~al.},\ }\href@noop {} {\bibfield  {journal} {\bibinfo  {journal} {Physical Review D}\ }\textbf {\bibinfo {volume} {103}},\ \bibinfo {pages} {122004} (\bibinfo {year} {2021})}\BibitemShut {NoStop}%
\bibitem [{\citenamefont {Ballmer}\ \emph {et~al.}(2024)\citenamefont {Ballmer}, \citenamefont {Barsotti}, \citenamefont {Kuns}, \citenamefont {Penn}, \citenamefont {Reid}, \citenamefont {Billingsley}, \citenamefont {Evans}, \citenamefont {Fejer}, \citenamefont {Harry},\ and\ \citenamefont {Menoni}}]{CEcoatingsWP}%
  \BibitemOpen
  \bibfield  {author} {\bibinfo {author} {\bibfnamefont {S.}~\bibnamefont {Ballmer}}, \bibinfo {author} {\bibfnamefont {L.}~\bibnamefont {Barsotti}}, \bibinfo {author} {\bibfnamefont {K.}~\bibnamefont {Kuns}}, \bibinfo {author} {\bibfnamefont {S.}~\bibnamefont {Penn}}, \bibinfo {author} {\bibfnamefont {S.}~\bibnamefont {Reid}}, \bibinfo {author} {\bibfnamefont {G.}~\bibnamefont {Billingsley}}, \bibinfo {author} {\bibfnamefont {M.}~\bibnamefont {Evans}}, \bibinfo {author} {\bibfnamefont {M.}~\bibnamefont {Fejer}}, \bibinfo {author} {\bibfnamefont {G.}~\bibnamefont {Harry}},\ and\ \bibinfo {author} {\bibfnamefont {C.}~\bibnamefont {Menoni}},\ }\href {https://dcc.cosmicexplorer.org/CE-T2400019/public} {\bibinfo {title} {The cosmic explorer observatory coating requirements and {R\&D}}},\ \bibinfo {howpublished} {Technical note (CE-T2400019)} (\bibinfo {year} {2024}),\ \bibinfo {note} {cE-T2400019-v4}\BibitemShut {NoStop}%
\bibitem [{\citenamefont {Abernathy}\ \emph {et~al.}(2011)\citenamefont {Abernathy}, \citenamefont {Reid}, \citenamefont {Chalkley}, \citenamefont {Bassiri}, \citenamefont {Martin}, \citenamefont {Evans}, \citenamefont {Fejer}, \citenamefont {Gretarsson}, \citenamefont {Harry}, \citenamefont {Hough} \emph {et~al.}}]{abernathy2011cryogenic}%
  \BibitemOpen
  \bibfield  {author} {\bibinfo {author} {\bibfnamefont {M.}~\bibnamefont {Abernathy}}, \bibinfo {author} {\bibfnamefont {S.}~\bibnamefont {Reid}}, \bibinfo {author} {\bibfnamefont {E.}~\bibnamefont {Chalkley}}, \bibinfo {author} {\bibfnamefont {R.}~\bibnamefont {Bassiri}}, \bibinfo {author} {\bibfnamefont {I.}~\bibnamefont {Martin}}, \bibinfo {author} {\bibfnamefont {K.}~\bibnamefont {Evans}}, \bibinfo {author} {\bibfnamefont {M.}~\bibnamefont {Fejer}}, \bibinfo {author} {\bibfnamefont {A.}~\bibnamefont {Gretarsson}}, \bibinfo {author} {\bibfnamefont {G.}~\bibnamefont {Harry}}, \bibinfo {author} {\bibfnamefont {J.}~\bibnamefont {Hough}}, \emph {et~al.},\ }\href@noop {} {\bibfield  {journal} {\bibinfo  {journal} {Classical and Quantum Gravity}\ }\textbf {\bibinfo {volume} {28}},\ \bibinfo {pages} {195017} (\bibinfo {year} {2011})}\BibitemShut {NoStop}%
\bibitem [{\citenamefont {Bassiri}(2011)}]{bassiri2011atomic}%
  \BibitemOpen
  \bibfield  {author} {\bibinfo {author} {\bibfnamefont {R.}~\bibnamefont {Bassiri}},\ }\emph {\bibinfo {title} {The atomic structure and properties of mirror coatings for use in gravitational wave detectors}},\ \href@noop {} {Ph.D. thesis},\ \bibinfo  {school} {University of Glasgow} (\bibinfo {year} {2011})\BibitemShut {NoStop}%
\bibitem [{\citenamefont {Hill}\ \emph {et~al.}(2008)\citenamefont {Hill}, \citenamefont {Bartynski}, \citenamefont {Nguyen}, \citenamefont {Davydov}, \citenamefont {Chandler-Horowitz},\ and\ \citenamefont {Frank}}]{hill2008relationship}%
  \BibitemOpen
  \bibfield  {author} {\bibinfo {author} {\bibfnamefont {D.~H.}\ \bibnamefont {Hill}}, \bibinfo {author} {\bibfnamefont {R.~A.}\ \bibnamefont {Bartynski}}, \bibinfo {author} {\bibfnamefont {N.~V.}\ \bibnamefont {Nguyen}}, \bibinfo {author} {\bibfnamefont {A.~C.}\ \bibnamefont {Davydov}}, \bibinfo {author} {\bibfnamefont {D.}~\bibnamefont {Chandler-Horowitz}},\ and\ \bibinfo {author} {\bibfnamefont {M.~M.}\ \bibnamefont {Frank}},\ }\href {https://doi.org/10.1063/1.2909442} {\bibfield  {journal} {\bibinfo  {journal} {Journal of Applied Physics}\ }\textbf {\bibinfo {volume} {103}},\ \bibinfo {pages} {093712} (\bibinfo {year} {2008})}\BibitemShut {NoStop}%
\bibitem [{\citenamefont {Ushakov}\ \emph {et~al.}(2004)\citenamefont {Ushakov}, \citenamefont {Navrotsky}, \citenamefont {Yang}, \citenamefont {Stemmer}, \citenamefont {Kukli}, \citenamefont {Ritala}, \citenamefont {Leskel{\"a}}, \citenamefont {Fejes}, \citenamefont {Demkov}, \citenamefont {Wang} \emph {et~al.}}]{ushakov2004crystallization}%
  \BibitemOpen
  \bibfield  {author} {\bibinfo {author} {\bibfnamefont {S.~V.}\ \bibnamefont {Ushakov}}, \bibinfo {author} {\bibfnamefont {A.}~\bibnamefont {Navrotsky}}, \bibinfo {author} {\bibfnamefont {Y.}~\bibnamefont {Yang}}, \bibinfo {author} {\bibfnamefont {S.}~\bibnamefont {Stemmer}}, \bibinfo {author} {\bibfnamefont {K.}~\bibnamefont {Kukli}}, \bibinfo {author} {\bibfnamefont {M.}~\bibnamefont {Ritala}}, \bibinfo {author} {\bibfnamefont {M.}~\bibnamefont {Leskel{\"a}}}, \bibinfo {author} {\bibfnamefont {P.}~\bibnamefont {Fejes}}, \bibinfo {author} {\bibfnamefont {A.}~\bibnamefont {Demkov}}, \bibinfo {author} {\bibfnamefont {C.}~\bibnamefont {Wang}}, \emph {et~al.},\ }\href@noop {} {\bibfield  {journal} {\bibinfo  {journal} {physica status solidi (b)}\ }\textbf {\bibinfo {volume} {241}},\ \bibinfo {pages} {2268} (\bibinfo {year} {2004})}\BibitemShut {NoStop}%
\bibitem [{\citenamefont {Afify}\ \emph {et~al.}(2006{\natexlab{a}})\citenamefont {Afify}, \citenamefont {Dalba}, \citenamefont {Koppolu}, \citenamefont {Armellini}, \citenamefont {Jestin},\ and\ \citenamefont {Rocca}}]{afify2006xrd}%
  \BibitemOpen
  \bibfield  {author} {\bibinfo {author} {\bibfnamefont {N.~D.}\ \bibnamefont {Afify}}, \bibinfo {author} {\bibfnamefont {G.}~\bibnamefont {Dalba}}, \bibinfo {author} {\bibfnamefont {U.~M.~K.}\ \bibnamefont {Koppolu}}, \bibinfo {author} {\bibfnamefont {C.}~\bibnamefont {Armellini}}, \bibinfo {author} {\bibfnamefont {Y.}~\bibnamefont {Jestin}},\ and\ \bibinfo {author} {\bibfnamefont {F.}~\bibnamefont {Rocca}},\ }\href {https://doi.org/10.1016/j.mssp.2006.10.021} {\bibfield  {journal} {\bibinfo  {journal} {Materials Science in Semiconductor Processing}\ }\textbf {\bibinfo {volume} {9}},\ \bibinfo {pages} {1043} (\bibinfo {year} {2006}{\natexlab{a}})}\BibitemShut {NoStop}%
\bibitem [{\citenamefont {Craig}\ \emph {et~al.}(2019)\citenamefont {Craig}, \citenamefont {Steinlechner}, \citenamefont {Murray} \emph {et~al.}}]{Craig2019PRL}%
  \BibitemOpen
  \bibfield  {author} {\bibinfo {author} {\bibfnamefont {K.}~\bibnamefont {Craig}}, \bibinfo {author} {\bibfnamefont {J.}~\bibnamefont {Steinlechner}}, \bibinfo {author} {\bibfnamefont {P.~G.}\ \bibnamefont {Murray}}, \emph {et~al.},\ }\href {https://doi.org/10.1103/physrevlett.122.231102} {\bibfield  {journal} {\bibinfo  {journal} {Phys. Rev. Lett.}\ }\textbf {\bibinfo {volume} {122}},\ \bibinfo {pages} {231102} (\bibinfo {year} {2019})}\BibitemShut {NoStop}%
\bibitem [{\citenamefont {Prasai}\ \emph {et~al.}(2019)\citenamefont {Prasai}, \citenamefont {Jiang}, \citenamefont {Mishkin}, \citenamefont {Shyam}, \citenamefont {Angelova}, \citenamefont {Birney}, \citenamefont {Drabold}, \citenamefont {Fazio}, \citenamefont {Gustafson}, \citenamefont {Harry} \emph {et~al.}}]{prasai2019high}%
  \BibitemOpen
  \bibfield  {author} {\bibinfo {author} {\bibfnamefont {K.}~\bibnamefont {Prasai}}, \bibinfo {author} {\bibfnamefont {J.}~\bibnamefont {Jiang}}, \bibinfo {author} {\bibfnamefont {A.}~\bibnamefont {Mishkin}}, \bibinfo {author} {\bibfnamefont {B.}~\bibnamefont {Shyam}}, \bibinfo {author} {\bibfnamefont {S.}~\bibnamefont {Angelova}}, \bibinfo {author} {\bibfnamefont {R.}~\bibnamefont {Birney}}, \bibinfo {author} {\bibfnamefont {D.}~\bibnamefont {Drabold}}, \bibinfo {author} {\bibfnamefont {M.}~\bibnamefont {Fazio}}, \bibinfo {author} {\bibfnamefont {E.}~\bibnamefont {Gustafson}}, \bibinfo {author} {\bibfnamefont {G.}~\bibnamefont {Harry}}, \emph {et~al.},\ }\href@noop {} {\bibfield  {journal} {\bibinfo  {journal} {Physical Review Letters}\ }\textbf {\bibinfo {volume} {123}},\ \bibinfo {pages} {045501} (\bibinfo {year} {2019})}\BibitemShut {NoStop}%
\bibitem [{\citenamefont {Prasai}\ \emph {et~al.}(2026)\citenamefont {Prasai}, \citenamefont {Jiang}, \citenamefont {Mishkin} \emph {et~al.}}]{prasai-submitted}%
  \BibitemOpen
  \bibfield  {author} {\bibinfo {author} {\bibfnamefont {K.}~\bibnamefont {Prasai}}, \bibinfo {author} {\bibfnamefont {J.}~\bibnamefont {Jiang}}, \bibinfo {author} {\bibfnamefont {A.}~\bibnamefont {Mishkin}}, \emph {et~al.},\ }\href@noop {} {\bibinfo {title} {Atomic structure of amorphous optical coatings of {TiO$_2$}-doped {GeO$_2$}~by grazing incidence total {X}-ray scattering measurements}} (\bibinfo {year} {2026}),\ \bibinfo {note} {submitted}\BibitemShut {NoStop}%
\bibitem [{\citenamefont {Puosi}\ \emph {et~al.}(2020)\citenamefont {Puosi}, \citenamefont {Fidecaro}, \citenamefont {Capaccioli}, \citenamefont {Pisignano},\ and\ \citenamefont {Leporini}}]{puosi2020nonlocal}%
  \BibitemOpen
  \bibfield  {author} {\bibinfo {author} {\bibfnamefont {F.}~\bibnamefont {Puosi}}, \bibinfo {author} {\bibfnamefont {F.}~\bibnamefont {Fidecaro}}, \bibinfo {author} {\bibfnamefont {S.}~\bibnamefont {Capaccioli}}, \bibinfo {author} {\bibfnamefont {D.}~\bibnamefont {Pisignano}},\ and\ \bibinfo {author} {\bibfnamefont {D.}~\bibnamefont {Leporini}},\ }\href {https://doi.org/10.1016/j.actamat.2020.09.054} {\bibfield  {journal} {\bibinfo  {journal} {Acta Materialia}\ }\textbf {\bibinfo {volume} {201}},\ \bibinfo {pages} {1} (\bibinfo {year} {2020})}\BibitemShut {NoStop}%
\bibitem [{\citenamefont {Damart}\ and\ \citenamefont {Rodney}(2018)}]{damart2018atomistic}%
  \BibitemOpen
  \bibfield  {author} {\bibinfo {author} {\bibfnamefont {T.}~\bibnamefont {Damart}}\ and\ \bibinfo {author} {\bibfnamefont {D.}~\bibnamefont {Rodney}},\ }\href@noop {} {\bibfield  {journal} {\bibinfo  {journal} {Physical Review B}\ }\textbf {\bibinfo {volume} {97}},\ \bibinfo {pages} {014201} (\bibinfo {year} {2018})}\BibitemShut {NoStop}%
\bibitem [{\citenamefont {Vajente}\ \emph {et~al.}(2018)\citenamefont {Vajente}, \citenamefont {Birney}, \citenamefont {Ananyeva}, \citenamefont {Angelova}, \citenamefont {Asselin}, \citenamefont {Baloukas}, \citenamefont {Bassiri}, \citenamefont {Billingsley}, \citenamefont {Fejer}, \citenamefont {Gibson} \emph {et~al.}}]{vajente2018effect}%
  \BibitemOpen
  \bibfield  {author} {\bibinfo {author} {\bibfnamefont {G.}~\bibnamefont {Vajente}}, \bibinfo {author} {\bibfnamefont {R.}~\bibnamefont {Birney}}, \bibinfo {author} {\bibfnamefont {A.}~\bibnamefont {Ananyeva}}, \bibinfo {author} {\bibfnamefont {S.}~\bibnamefont {Angelova}}, \bibinfo {author} {\bibfnamefont {R.}~\bibnamefont {Asselin}}, \bibinfo {author} {\bibfnamefont {B.}~\bibnamefont {Baloukas}}, \bibinfo {author} {\bibfnamefont {R.}~\bibnamefont {Bassiri}}, \bibinfo {author} {\bibfnamefont {G.}~\bibnamefont {Billingsley}}, \bibinfo {author} {\bibfnamefont {M.}~\bibnamefont {Fejer}}, \bibinfo {author} {\bibfnamefont {D.}~\bibnamefont {Gibson}}, \emph {et~al.},\ }\href@noop {} {\bibfield  {journal} {\bibinfo  {journal} {Classical and Quantum Gravity}\ }\textbf {\bibinfo {volume} {35}},\ \bibinfo {pages} {075001} (\bibinfo {year} {2018})}\BibitemShut {NoStop}%
\bibitem [{\citenamefont {Fazio}\ \emph {et~al.}(2022)\citenamefont {Fazio}, \citenamefont {Vajente}, \citenamefont {Yang}, \citenamefont {Ananyeva},\ and\ \citenamefont {Menoni}}]{fazio2022comprehensive}%
  \BibitemOpen
  \bibfield  {author} {\bibinfo {author} {\bibfnamefont {M.~A.}\ \bibnamefont {Fazio}}, \bibinfo {author} {\bibfnamefont {G.}~\bibnamefont {Vajente}}, \bibinfo {author} {\bibfnamefont {L.}~\bibnamefont {Yang}}, \bibinfo {author} {\bibfnamefont {A.}~\bibnamefont {Ananyeva}},\ and\ \bibinfo {author} {\bibfnamefont {C.~S.}\ \bibnamefont {Menoni}},\ }\href@noop {} {\bibfield  {journal} {\bibinfo  {journal} {Physical Review D}\ }\textbf {\bibinfo {volume} {105}},\ \bibinfo {pages} {102008} (\bibinfo {year} {2022})}\BibitemShut {NoStop}%
\bibitem [{\citenamefont {Cho}\ \emph {et~al.}(2008)\citenamefont {Cho}, \citenamefont {Park}, \citenamefont {Na}, \citenamefont {Kim},\ and\ \citenamefont {Hwang}}]{Cho2008HfO2EXAFS}%
  \BibitemOpen
  \bibfield  {author} {\bibinfo {author} {\bibfnamefont {D.-Y.}\ \bibnamefont {Cho}}, \bibinfo {author} {\bibfnamefont {T.~J.}\ \bibnamefont {Park}}, \bibinfo {author} {\bibfnamefont {K.~D.}\ \bibnamefont {Na}}, \bibinfo {author} {\bibfnamefont {J.~H.}\ \bibnamefont {Kim}},\ and\ \bibinfo {author} {\bibfnamefont {C.~S.}\ \bibnamefont {Hwang}},\ }\href {https://doi.org/10.1103/PhysRevB.78.132102} {\bibfield  {journal} {\bibinfo  {journal} {Physical Review B}\ }\textbf {\bibinfo {volume} {78}},\ \bibinfo {pages} {132102} (\bibinfo {year} {2008})}\BibitemShut {NoStop}%
\bibitem [{\citenamefont {Haque}\ \emph {et~al.}(2016)\citenamefont {Haque}, \citenamefont {Nayak}, \citenamefont {Bhattacharyya}, \citenamefont {Jha},\ and\ \citenamefont {Sahoo}}]{Haque2016HfO2EXAFS}%
  \BibitemOpen
  \bibfield  {author} {\bibinfo {author} {\bibfnamefont {S.~M.}\ \bibnamefont {Haque}}, \bibinfo {author} {\bibfnamefont {C.}~\bibnamefont {Nayak}}, \bibinfo {author} {\bibfnamefont {D.}~\bibnamefont {Bhattacharyya}}, \bibinfo {author} {\bibfnamefont {S.~N.}\ \bibnamefont {Jha}},\ and\ \bibinfo {author} {\bibfnamefont {N.~K.}\ \bibnamefont {Sahoo}},\ }\href {https://doi.org/10.1364/AO.55.002175} {\bibfield  {journal} {\bibinfo  {journal} {Applied Optics}\ }\textbf {\bibinfo {volume} {55}},\ \bibinfo {pages} {2175} (\bibinfo {year} {2016})}\BibitemShut {NoStop}%
\bibitem [{\citenamefont {Afify}\ \emph {et~al.}(2006{\natexlab{b}})\citenamefont {Afify}, \citenamefont {Dalba}, \citenamefont {Koppolu}, \citenamefont {Armellini}, \citenamefont {Jestin},\ and\ \citenamefont {Rocca}}]{Afify2006SiO2HfO2EXAFS}%
  \BibitemOpen
  \bibfield  {author} {\bibinfo {author} {\bibfnamefont {N.~D.}\ \bibnamefont {Afify}}, \bibinfo {author} {\bibfnamefont {G.}~\bibnamefont {Dalba}}, \bibinfo {author} {\bibfnamefont {U.~M.~K.}\ \bibnamefont {Koppolu}}, \bibinfo {author} {\bibfnamefont {C.}~\bibnamefont {Armellini}}, \bibinfo {author} {\bibfnamefont {Y.}~\bibnamefont {Jestin}},\ and\ \bibinfo {author} {\bibfnamefont {F.}~\bibnamefont {Rocca}},\ }\href {https://doi.org/10.1016/j.mssp.2006.10.021} {\bibfield  {journal} {\bibinfo  {journal} {Materials Science in Semiconductor Processing}\ }\textbf {\bibinfo {volume} {9}},\ \bibinfo {pages} {1043} (\bibinfo {year} {2006}{\natexlab{b}})}\BibitemShut {NoStop}%
\bibitem [{\citenamefont {Liu}\ \emph {et~al.}(2024)\citenamefont {Liu}, \citenamefont {Zaheer}, \citenamefont {Carrillo},\ and\ \citenamefont {Banerjee}}]{Liu2024HfO2Orthorhombic}%
  \BibitemOpen
  \bibfield  {author} {\bibinfo {author} {\bibfnamefont {G.-W.}\ \bibnamefont {Liu}}, \bibinfo {author} {\bibfnamefont {W.}~\bibnamefont {Zaheer}}, \bibinfo {author} {\bibfnamefont {L.}~\bibnamefont {Carrillo}},\ and\ \bibinfo {author} {\bibfnamefont {S.}~\bibnamefont {Banerjee}},\ }\href {https://doi.org/10.1016/j.xcrp.2024.101818} {\bibfield  {journal} {\bibinfo  {journal} {Cell Reports Physical Science}\ }\textbf {\bibinfo {volume} {5}},\ \bibinfo {pages} {101818} (\bibinfo {year} {2024})}\BibitemShut {NoStop}%
\bibitem [{\citenamefont {Ozkendir}\ \emph {et~al.}(2025)\citenamefont {Ozkendir}, \citenamefont {Cengiz}, \citenamefont {Kanmaz}, \citenamefont {Gunaydin}, \citenamefont {Apayd{\i}n},\ and\ \citenamefont {Harfouche}}]{Ozkendir2025HfO2XAFS}%
  \BibitemOpen
  \bibfield  {author} {\bibinfo {author} {\bibfnamefont {O.~M.}\ \bibnamefont {Ozkendir}}, \bibinfo {author} {\bibfnamefont {E.}~\bibnamefont {Cengiz}}, \bibinfo {author} {\bibfnamefont {\.{I}.}~\bibnamefont {Kanmaz}}, \bibinfo {author} {\bibfnamefont {S.}~\bibnamefont {Gunaydin}}, \bibinfo {author} {\bibfnamefont {G.}~\bibnamefont {Apayd{\i}n}},\ and\ \bibinfo {author} {\bibfnamefont {M.}~\bibnamefont {Harfouche}},\ }\href {https://doi.org/10.1007/s11664-025-12205-x} {\bibfield  {journal} {\bibinfo  {journal} {Journal of Electronic Materials}\ }\textbf {\bibinfo {volume} {54}},\ \bibinfo {pages} {10511} (\bibinfo {year} {2025})}\BibitemShut {NoStop}%
\bibitem [{\citenamefont {Nguyen}\ \emph {et~al.}(2019{\natexlab{a}})\citenamefont {Nguyen}, \citenamefont {Le},\ and\ \citenamefont {Nguyen}}]{Nguyen2019AmorphousHfO2}%
  \BibitemOpen
  \bibfield  {author} {\bibinfo {author} {\bibfnamefont {T.~H.}\ \bibnamefont {Nguyen}}, \bibinfo {author} {\bibfnamefont {V.~V.}\ \bibnamefont {Le}},\ and\ \bibinfo {author} {\bibfnamefont {T.~N.}\ \bibnamefont {Nguyen}},\ }\href {https://doi.org/10.1016/j.vacuum.2018.12.028} {\bibfield  {journal} {\bibinfo  {journal} {Vacuum}\ }\textbf {\bibinfo {volume} {161}},\ \bibinfo {pages} {251} (\bibinfo {year} {2019}{\natexlab{a}})}\BibitemShut {NoStop}%
\bibitem [{\citenamefont {Chen}\ and\ \citenamefont {Kuo}(2011)}]{Chen2011AmorphousHfO2}%
  \BibitemOpen
  \bibfield  {author} {\bibinfo {author} {\bibfnamefont {T.-J.}\ \bibnamefont {Chen}}\ and\ \bibinfo {author} {\bibfnamefont {C.-L.}\ \bibnamefont {Kuo}},\ }\href {https://doi.org/10.1063/1.3636362} {\bibfield  {journal} {\bibinfo  {journal} {Journal of Applied Physics}\ }\textbf {\bibinfo {volume} {110}},\ \bibinfo {pages} {064105} (\bibinfo {year} {2011})}\BibitemShut {NoStop}%
\bibitem [{\citenamefont {Kumar}\ \emph {et~al.}(2021)\citenamefont {Kumar}, \citenamefont {Singh},\ and\ \citenamefont {Kumar}}]{Kumar2021HfO2Optoelectronic}%
  \BibitemOpen
  \bibfield  {author} {\bibinfo {author} {\bibfnamefont {M.}~\bibnamefont {Kumar}}, \bibinfo {author} {\bibfnamefont {R.~P.}\ \bibnamefont {Singh}},\ and\ \bibinfo {author} {\bibfnamefont {A.}~\bibnamefont {Kumar}},\ }\href {https://doi.org/10.1016/j.ijleo.2020.165937} {\bibfield  {journal} {\bibinfo  {journal} {Optik}\ }\textbf {\bibinfo {volume} {226}},\ \bibinfo {pages} {165937} (\bibinfo {year} {2021})}\BibitemShut {NoStop}%
\bibitem [{\citenamefont {Perevalov}\ \emph {et~al.}(2007)\citenamefont {Perevalov}, \citenamefont {Gritsenko}, \citenamefont {Erenburg}, \citenamefont {Badalyan}, \citenamefont {Wong},\ and\ \citenamefont {Kim}}]{Perevalov2007HfO2}%
  \BibitemOpen
  \bibfield  {author} {\bibinfo {author} {\bibfnamefont {T.~V.}\ \bibnamefont {Perevalov}}, \bibinfo {author} {\bibfnamefont {V.~A.}\ \bibnamefont {Gritsenko}}, \bibinfo {author} {\bibfnamefont {S.~B.}\ \bibnamefont {Erenburg}}, \bibinfo {author} {\bibfnamefont {A.~M.}\ \bibnamefont {Badalyan}}, \bibinfo {author} {\bibfnamefont {H.}~\bibnamefont {Wong}},\ and\ \bibinfo {author} {\bibfnamefont {C.~W.}\ \bibnamefont {Kim}},\ }\href {https://doi.org/10.1063/1.2464184} {\bibfield  {journal} {\bibinfo  {journal} {Journal of Applied Physics}\ }\textbf {\bibinfo {volume} {101}},\ \bibinfo {pages} {053704} (\bibinfo {year} {2007})}\BibitemShut {NoStop}%
\bibitem [{\citenamefont {Zhao}\ and\ \citenamefont {Vanderbilt}(2002)}]{Zhao2002HfO2}%
  \BibitemOpen
  \bibfield  {author} {\bibinfo {author} {\bibfnamefont {X.}~\bibnamefont {Zhao}}\ and\ \bibinfo {author} {\bibfnamefont {D.}~\bibnamefont {Vanderbilt}},\ }\href {https://doi.org/10.1103/PhysRevB.65.233106} {\bibfield  {journal} {\bibinfo  {journal} {Physical Review B}\ }\textbf {\bibinfo {volume} {65}},\ \bibinfo {pages} {233106} (\bibinfo {year} {2002})}\BibitemShut {NoStop}%
\bibitem [{\citenamefont {Gallington}\ \emph {et~al.}(2017)\citenamefont {Gallington}, \citenamefont {Ghadar}, \citenamefont {Skinner}, \citenamefont {Weber}, \citenamefont {Ushakov}, \citenamefont {Navrotsky}, \citenamefont {Vazquez-Mayagoitia}, \citenamefont {Neuefeind}, \citenamefont {Stan}, \citenamefont {Low},\ and\ \citenamefont {Benmore}}]{gallington2017structure}%
  \BibitemOpen
  \bibfield  {author} {\bibinfo {author} {\bibfnamefont {L.~C.}\ \bibnamefont {Gallington}}, \bibinfo {author} {\bibfnamefont {Y.}~\bibnamefont {Ghadar}}, \bibinfo {author} {\bibfnamefont {L.~B.}\ \bibnamefont {Skinner}}, \bibinfo {author} {\bibfnamefont {J.~K.~R.}\ \bibnamefont {Weber}}, \bibinfo {author} {\bibfnamefont {S.~V.}\ \bibnamefont {Ushakov}}, \bibinfo {author} {\bibfnamefont {A.}~\bibnamefont {Navrotsky}}, \bibinfo {author} {\bibfnamefont {A.}~\bibnamefont {Vazquez-Mayagoitia}}, \bibinfo {author} {\bibfnamefont {J.~C.}\ \bibnamefont {Neuefeind}}, \bibinfo {author} {\bibfnamefont {M.}~\bibnamefont {Stan}}, \bibinfo {author} {\bibfnamefont {J.~J.}\ \bibnamefont {Low}},\ and\ \bibinfo {author} {\bibfnamefont {C.~J.}\ \bibnamefont {Benmore}},\ }\href {https://doi.org/10.3390/ma10111290} {\bibfield  {journal} {\bibinfo  {journal} {Materials}\ }\textbf {\bibinfo {volume} {10}},\ \bibinfo {pages} {1290} (\bibinfo {year} {2017})}\BibitemShut {NoStop}%
\bibitem [{\citenamefont {Nguyen}\ \emph {et~al.}(2019{\natexlab{b}})\citenamefont {Nguyen}, \citenamefont {Le},\ and\ \citenamefont {Nguyen}}]{nguyen2019molecular}%
  \BibitemOpen
  \bibfield  {author} {\bibinfo {author} {\bibfnamefont {T.~H.}\ \bibnamefont {Nguyen}}, \bibinfo {author} {\bibfnamefont {V.~V.}\ \bibnamefont {Le}},\ and\ \bibinfo {author} {\bibfnamefont {T.~N.}\ \bibnamefont {Nguyen}},\ }\href {https://doi.org/10.1016/j.vacuum.2018.12.028} {\bibfield  {journal} {\bibinfo  {journal} {Vacuum}\ }\textbf {\bibinfo {volume} {161}},\ \bibinfo {pages} {251} (\bibinfo {year} {2019}{\natexlab{b}})}\BibitemShut {NoStop}%
\bibitem [{\citenamefont {Jensen}\ \emph {et~al.}(2015)\citenamefont {Jensen}, \citenamefont {Blichfeld}, \citenamefont {Bauers}, \citenamefont {Wood}, \citenamefont {Dooryh{\'e}e}, \citenamefont {Johnson}, \citenamefont {Iversen},\ and\ \citenamefont {Billinge}}]{jensen2015demonstration}%
  \BibitemOpen
  \bibfield  {author} {\bibinfo {author} {\bibfnamefont {K.}~\bibnamefont {Jensen}}, \bibinfo {author} {\bibfnamefont {A.~B.}\ \bibnamefont {Blichfeld}}, \bibinfo {author} {\bibfnamefont {S.~R.}\ \bibnamefont {Bauers}}, \bibinfo {author} {\bibfnamefont {S.~R.}\ \bibnamefont {Wood}}, \bibinfo {author} {\bibfnamefont {E.}~\bibnamefont {Dooryh{\'e}e}}, \bibinfo {author} {\bibfnamefont {D.~C.}\ \bibnamefont {Johnson}}, \bibinfo {author} {\bibfnamefont {B.~B.}\ \bibnamefont {Iversen}},\ and\ \bibinfo {author} {\bibfnamefont {S.~J.}\ \bibnamefont {Billinge}},\ }\href@noop {} {\bibfield  {journal} {\bibinfo  {journal} {IUCrJ}\ }\textbf {\bibinfo {volume} {2}},\ \bibinfo {pages} {481} (\bibinfo {year} {2015})}\BibitemShut {NoStop}%
\bibitem [{\citenamefont {Shyam}\ \emph {et~al.}(2016)\citenamefont {Shyam}, \citenamefont {Stone}, \citenamefont {Bassiri}, \citenamefont {Fejer}, \citenamefont {Toney},\ and\ \citenamefont {Mehta}}]{shyam2016measurement}%
  \BibitemOpen
  \bibfield  {author} {\bibinfo {author} {\bibfnamefont {B.}~\bibnamefont {Shyam}}, \bibinfo {author} {\bibfnamefont {K.~H.}\ \bibnamefont {Stone}}, \bibinfo {author} {\bibfnamefont {R.}~\bibnamefont {Bassiri}}, \bibinfo {author} {\bibfnamefont {M.~M.}\ \bibnamefont {Fejer}}, \bibinfo {author} {\bibfnamefont {M.~F.}\ \bibnamefont {Toney}},\ and\ \bibinfo {author} {\bibfnamefont {A.}~\bibnamefont {Mehta}},\ }\href@noop {} {\bibfield  {journal} {\bibinfo  {journal} {Scientific Reports}\ }\textbf {\bibinfo {volume} {6}},\ \bibinfo {pages} {1} (\bibinfo {year} {2016})}\BibitemShut {NoStop}%
\bibitem [{\citenamefont {Qiu}\ \emph {et~al.}(2004)\citenamefont {Qiu}, \citenamefont {Thompson},\ and\ \citenamefont {Billinge}}]{qiu2004pdfgetx2}%
  \BibitemOpen
  \bibfield  {author} {\bibinfo {author} {\bibfnamefont {X.}~\bibnamefont {Qiu}}, \bibinfo {author} {\bibfnamefont {J.~W.}\ \bibnamefont {Thompson}},\ and\ \bibinfo {author} {\bibfnamefont {S.~J.}\ \bibnamefont {Billinge}},\ }\href@noop {} {\bibfield  {journal} {\bibinfo  {journal} {Journal of Applied Crystallography}\ }\textbf {\bibinfo {volume} {37}},\ \bibinfo {pages} {678} (\bibinfo {year} {2004})}\BibitemShut {NoStop}%
\bibitem [{\citenamefont {Birch}(1947)}]{birch1947finite}%
  \BibitemOpen
  \bibfield  {author} {\bibinfo {author} {\bibfnamefont {F.}~\bibnamefont {Birch}},\ }\href {https://doi.org/10.1103/PhysRev.71.809} {\bibfield  {journal} {\bibinfo  {journal} {Physical Review}\ }\textbf {\bibinfo {volume} {71}},\ \bibinfo {pages} {809} (\bibinfo {year} {1947})}\BibitemShut {NoStop}%
\bibitem [{\citenamefont {Modreanu}\ \emph {et~al.}(2005)\citenamefont {Modreanu}, \citenamefont {Sancho-Parramon}, \citenamefont {O’Connell}, \citenamefont {Justice}, \citenamefont {Durand},\ and\ \citenamefont {Servet}}]{modreanu2005solid}%
  \BibitemOpen
  \bibfield  {author} {\bibinfo {author} {\bibfnamefont {M.}~\bibnamefont {Modreanu}}, \bibinfo {author} {\bibfnamefont {J.}~\bibnamefont {Sancho-Parramon}}, \bibinfo {author} {\bibfnamefont {D.}~\bibnamefont {O’Connell}}, \bibinfo {author} {\bibfnamefont {J.}~\bibnamefont {Justice}}, \bibinfo {author} {\bibfnamefont {O.}~\bibnamefont {Durand}},\ and\ \bibinfo {author} {\bibfnamefont {B.}~\bibnamefont {Servet}},\ }\href@noop {} {\bibfield  {journal} {\bibinfo  {journal} {Materials Science and Engineering: B}\ }\textbf {\bibinfo {volume} {118}},\ \bibinfo {pages} {127} (\bibinfo {year} {2005})}\BibitemShut {NoStop}%
\bibitem [{\citenamefont {Puurunen}\ \emph {et~al.}(2005)\citenamefont {Puurunen}, \citenamefont {Delabie}, \citenamefont {Van~Elshocht}, \citenamefont {Caymax}, \citenamefont {Green}, \citenamefont {Brijs}, \citenamefont {Richard}, \citenamefont {Bender}, \citenamefont {Conard}, \citenamefont {Hoflijk} \emph {et~al.}}]{puurunen2005hafnium}%
  \BibitemOpen
  \bibfield  {author} {\bibinfo {author} {\bibfnamefont {R.~L.}\ \bibnamefont {Puurunen}}, \bibinfo {author} {\bibfnamefont {A.}~\bibnamefont {Delabie}}, \bibinfo {author} {\bibfnamefont {S.}~\bibnamefont {Van~Elshocht}}, \bibinfo {author} {\bibfnamefont {M.}~\bibnamefont {Caymax}}, \bibinfo {author} {\bibfnamefont {M.~L.}\ \bibnamefont {Green}}, \bibinfo {author} {\bibfnamefont {B.}~\bibnamefont {Brijs}}, \bibinfo {author} {\bibfnamefont {O.}~\bibnamefont {Richard}}, \bibinfo {author} {\bibfnamefont {H.}~\bibnamefont {Bender}}, \bibinfo {author} {\bibfnamefont {T.}~\bibnamefont {Conard}}, \bibinfo {author} {\bibfnamefont {I.}~\bibnamefont {Hoflijk}}, \emph {et~al.},\ }\href@noop {} {\bibfield  {journal} {\bibinfo  {journal} {Applied Physics Letters}\ }\textbf {\bibinfo {volume} {86}} (\bibinfo {year} {2005})}\BibitemShut {NoStop}%
\bibitem [{\citenamefont {Batatia}\ \emph {et~al.}(2022)\citenamefont {Batatia}, \citenamefont {Kov{\'a}cs}, \citenamefont {Simm}, \citenamefont {Ortner},\ and\ \citenamefont {Cs{\'a}nyi}}]{batatia2022mace}%
  \BibitemOpen
  \bibfield  {author} {\bibinfo {author} {\bibfnamefont {I.}~\bibnamefont {Batatia}}, \bibinfo {author} {\bibfnamefont {D.~P.}\ \bibnamefont {Kov{\'a}cs}}, \bibinfo {author} {\bibfnamefont {G.~N.~C.}\ \bibnamefont {Simm}}, \bibinfo {author} {\bibfnamefont {C.}~\bibnamefont {Ortner}},\ and\ \bibinfo {author} {\bibfnamefont {G.}~\bibnamefont {Cs{\'a}nyi}},\ }\href@noop {} {\bibfield  {journal} {\bibinfo  {journal} {arXiv preprint arXiv:2206.07697}\ } (\bibinfo {year} {2022})}\BibitemShut {NoStop}%
\bibitem [{\citenamefont {Kresse}\ and\ \citenamefont {Furthm{\"u}ller}(1996)}]{kresse1996efficient}%
  \BibitemOpen
  \bibfield  {author} {\bibinfo {author} {\bibfnamefont {G.}~\bibnamefont {Kresse}}\ and\ \bibinfo {author} {\bibfnamefont {J.}~\bibnamefont {Furthm{\"u}ller}},\ }\href@noop {} {\bibfield  {journal} {\bibinfo  {journal} {Physical Review B}\ }\textbf {\bibinfo {volume} {54}},\ \bibinfo {pages} {11169} (\bibinfo {year} {1996})}\BibitemShut {NoStop}%
\bibitem [{\citenamefont {Perdew}\ \emph {et~al.}(1996)\citenamefont {Perdew}, \citenamefont {Burke},\ and\ \citenamefont {Ernzerhof}}]{perdew1996generalized}%
  \BibitemOpen
  \bibfield  {author} {\bibinfo {author} {\bibfnamefont {J.~P.}\ \bibnamefont {Perdew}}, \bibinfo {author} {\bibfnamefont {K.}~\bibnamefont {Burke}},\ and\ \bibinfo {author} {\bibfnamefont {M.}~\bibnamefont {Ernzerhof}},\ }\href@noop {} {\bibfield  {journal} {\bibinfo  {journal} {Physical Review Letters}\ }\textbf {\bibinfo {volume} {77}},\ \bibinfo {pages} {3865} (\bibinfo {year} {1996})}\BibitemShut {NoStop}%
\bibitem [{\citenamefont {Bl{\"o}chl}(1994)}]{blochl1994projector}%
  \BibitemOpen
  \bibfield  {author} {\bibinfo {author} {\bibfnamefont {P.~E.}\ \bibnamefont {Bl{\"o}chl}},\ }\href@noop {} {\bibfield  {journal} {\bibinfo  {journal} {Physical Review B}\ }\textbf {\bibinfo {volume} {50}},\ \bibinfo {pages} {17953} (\bibinfo {year} {1994})}\BibitemShut {NoStop}%
\bibitem [{\citenamefont {Drabold}(2009)}]{drabold2009topics}%
  \BibitemOpen
  \bibfield  {author} {\bibinfo {author} {\bibfnamefont {D.}~\bibnamefont {Drabold}},\ }\href@noop {} {\bibfield  {journal} {\bibinfo  {journal} {The European Physical Journal B}\ }\textbf {\bibinfo {volume} {68}},\ \bibinfo {pages} {1} (\bibinfo {year} {2009})}\BibitemShut {NoStop}%
\bibitem [{\citenamefont {Keen}\ and\ \citenamefont {McGreevy}(1990)}]{keen1990structural}%
  \BibitemOpen
  \bibfield  {author} {\bibinfo {author} {\bibfnamefont {D.}~\bibnamefont {Keen}}\ and\ \bibinfo {author} {\bibfnamefont {R.}~\bibnamefont {McGreevy}},\ }\href@noop {} {\bibfield  {journal} {\bibinfo  {journal} {Nature}\ }\textbf {\bibinfo {volume} {344}},\ \bibinfo {pages} {423} (\bibinfo {year} {1990})}\BibitemShut {NoStop}%
\bibitem [{\citenamefont {McGreevy}(2001)}]{mcgreevy2001reverse}%
  \BibitemOpen
  \bibfield  {author} {\bibinfo {author} {\bibfnamefont {R.~L.}\ \bibnamefont {McGreevy}},\ }\href@noop {} {\bibfield  {journal} {\bibinfo  {journal} {Journal of Physics: Condensed Matter}\ }\textbf {\bibinfo {volume} {13}},\ \bibinfo {pages} {R877} (\bibinfo {year} {2001})}\BibitemShut {NoStop}%
\bibitem [{\citenamefont {Pandey}\ \emph {et~al.}(2015)\citenamefont {Pandey}, \citenamefont {Biswas},\ and\ \citenamefont {Drabold}}]{pandey2015force}%
  \BibitemOpen
  \bibfield  {author} {\bibinfo {author} {\bibfnamefont {A.}~\bibnamefont {Pandey}}, \bibinfo {author} {\bibfnamefont {P.}~\bibnamefont {Biswas}},\ and\ \bibinfo {author} {\bibfnamefont {D.}~\bibnamefont {Drabold}},\ }\href@noop {} {\bibfield  {journal} {\bibinfo  {journal} {Physical Review B}\ }\textbf {\bibinfo {volume} {92}},\ \bibinfo {pages} {155205} (\bibinfo {year} {2015})}\BibitemShut {NoStop}%
\bibitem [{\citenamefont {Pandey}\ \emph {et~al.}(2016)\citenamefont {Pandey}, \citenamefont {Biswas},\ and\ \citenamefont {Drabold}}]{pandey2016inversion}%
  \BibitemOpen
  \bibfield  {author} {\bibinfo {author} {\bibfnamefont {A.}~\bibnamefont {Pandey}}, \bibinfo {author} {\bibfnamefont {P.}~\bibnamefont {Biswas}},\ and\ \bibinfo {author} {\bibfnamefont {D.~A.}\ \bibnamefont {Drabold}},\ }\href@noop {} {\bibfield  {journal} {\bibinfo  {journal} {Scientific Reports}\ }\textbf {\bibinfo {volume} {6}},\ \bibinfo {pages} {1} (\bibinfo {year} {2016})}\BibitemShut {NoStop}%
\bibitem [{\citenamefont {Bitzek}\ \emph {et~al.}(2006)\citenamefont {Bitzek}, \citenamefont {Koskinen}, \citenamefont {G{\"a}hler}, \citenamefont {Moseler},\ and\ \citenamefont {Gumbsch}}]{bitzek2006structural}%
  \BibitemOpen
  \bibfield  {author} {\bibinfo {author} {\bibfnamefont {E.}~\bibnamefont {Bitzek}}, \bibinfo {author} {\bibfnamefont {P.}~\bibnamefont {Koskinen}}, \bibinfo {author} {\bibfnamefont {F.}~\bibnamefont {G{\"a}hler}}, \bibinfo {author} {\bibfnamefont {M.}~\bibnamefont {Moseler}},\ and\ \bibinfo {author} {\bibfnamefont {P.}~\bibnamefont {Gumbsch}},\ }\href {https://doi.org/10.1103/PhysRevLett.97.170201} {\bibfield  {journal} {\bibinfo  {journal} {Physical Review Letters}\ }\textbf {\bibinfo {volume} {97}},\ \bibinfo {pages} {170201} (\bibinfo {year} {2006})}\BibitemShut {NoStop}%
\bibitem [{\citenamefont {Norberg}\ \emph {et~al.}(2009)\citenamefont {Norberg}, \citenamefont {Tucker},\ and\ \citenamefont {Hull}}]{norberg2009bond}%
  \BibitemOpen
  \bibfield  {author} {\bibinfo {author} {\bibfnamefont {S.~T.}\ \bibnamefont {Norberg}}, \bibinfo {author} {\bibfnamefont {M.~G.}\ \bibnamefont {Tucker}},\ and\ \bibinfo {author} {\bibfnamefont {S.}~\bibnamefont {Hull}},\ }\href@noop {} {\bibfield  {journal} {\bibinfo  {journal} {Applied Crystallography}\ }\textbf {\bibinfo {volume} {42}},\ \bibinfo {pages} {179} (\bibinfo {year} {2009})}\BibitemShut {NoStop}%
\bibitem [{\citenamefont {Keen}(2001)}]{keen2001comparison}%
  \BibitemOpen
  \bibfield  {author} {\bibinfo {author} {\bibfnamefont {D.~A.}\ \bibnamefont {Keen}},\ }\href@noop {} {\bibfield  {journal} {\bibinfo  {journal} {Applied Crystallography}\ }\textbf {\bibinfo {volume} {34}},\ \bibinfo {pages} {172} (\bibinfo {year} {2001})}\BibitemShut {NoStop}%
\bibitem [{\citenamefont {Elliott}(1991)}]{elliott1991medium}%
  \BibitemOpen
  \bibfield  {author} {\bibinfo {author} {\bibfnamefont {S.~R.}\ \bibnamefont {Elliott}},\ }\href@noop {} {\bibfield  {journal} {\bibinfo  {journal} {Nature}\ }\textbf {\bibinfo {volume} {354}},\ \bibinfo {pages} {445} (\bibinfo {year} {1991})}\BibitemShut {NoStop}%
\bibitem [{\citenamefont {Mishkin}\ \emph {et~al.}(2023)\citenamefont {Mishkin}, \citenamefont {Jiang}, \citenamefont {Zhang}, \citenamefont {Cheng}, \citenamefont {Prasai}, \citenamefont {Bassiri},\ and\ \citenamefont {Fejer}}]{mishkin2023hidden}%
  \BibitemOpen
  \bibfield  {author} {\bibinfo {author} {\bibfnamefont {A.}~\bibnamefont {Mishkin}}, \bibinfo {author} {\bibfnamefont {J.}~\bibnamefont {Jiang}}, \bibinfo {author} {\bibfnamefont {R.}~\bibnamefont {Zhang}}, \bibinfo {author} {\bibfnamefont {H.-P.}\ \bibnamefont {Cheng}}, \bibinfo {author} {\bibfnamefont {K.}~\bibnamefont {Prasai}}, \bibinfo {author} {\bibfnamefont {R.}~\bibnamefont {Bassiri}},\ and\ \bibinfo {author} {\bibfnamefont {M.}~\bibnamefont {Fejer}},\ }\href@noop {} {\bibfield  {journal} {\bibinfo  {journal} {Physical Review B}\ }\textbf {\bibinfo {volume} {108}},\ \bibinfo {pages} {054103} (\bibinfo {year} {2023})}\BibitemShut {NoStop}%
\bibitem [{\citenamefont {Prasai}\ \emph {et~al.}(2021)\citenamefont {Prasai}, \citenamefont {Bassiri}, \citenamefont {Cheng},\ and\ \citenamefont {Fejer}}]{prasai2021annealing}%
  \BibitemOpen
  \bibfield  {author} {\bibinfo {author} {\bibfnamefont {K.}~\bibnamefont {Prasai}}, \bibinfo {author} {\bibfnamefont {R.}~\bibnamefont {Bassiri}}, \bibinfo {author} {\bibfnamefont {H.-P.}\ \bibnamefont {Cheng}},\ and\ \bibinfo {author} {\bibfnamefont {M.~M.}\ \bibnamefont {Fejer}},\ }\href@noop {} {\bibfield  {journal} {\bibinfo  {journal} {physica status solidi (b)}\ }\textbf {\bibinfo {volume} {258}},\ \bibinfo {pages} {2000519} (\bibinfo {year} {2021})}\BibitemShut {NoStop}%
\end{thebibliography}%
\bibliographystyle{apsrev4-2}

\end{document}